\documentclass[reprint, superscriptaddress, nofootinbib, amsmath, amssymb, prl,]{revtex4-2}

\usepackage{graphicx}
\usepackage{dcolumn}
\usepackage{bm}
\usepackage{xcolor}
\usepackage{hyperref}
\hypersetup{
    colorlinks=true,
    linkcolor=blue,
    citecolor=blue,
    urlcolor=blue}
    
\begin{document}

\title{Weakly magnetized, Hall dominated plasma Couette flow}

\author{K.~ Flanagan}
\affiliation{Department of Physics, University of Wisconsin--Madison, 1150 University Avenue, Madison, WI 53706, USA}
\author{J.~ Milhone}
\affiliation{Department of Physics, University of Wisconsin--Madison, 1150 University Avenue, Madison, WI 53706, USA}
\author{J.~Egedal}
\affiliation{Department of Physics, University of Wisconsin--Madison, 1150 University Avenue, Madison, WI 53706, USA}
\author{D.~ Endrizzi}
\affiliation{Department of Physics, University of Wisconsin--Madison, 1150 University Avenue, Madison, WI 53706, USA}
\author{J.~ Olson}
\affiliation{Department of Physics, University of Wisconsin--Madison, 1150 University Avenue, Madison, WI 53706, USA}
\author{E.~ E.~Peterson}
\affiliation{Plasma Science and Fusion Center, Massachusetts Institute of Technology, 77 Massachusetts Avenue, NW 17 Cambridge, MA 02139}
\author{R.~Sassella}
\affiliation{Department of Physics, University of Wisconsin--Madison, 1150 University Avenue, Madison, WI 53706, USA}
\author{C.~ B.~Forest}
\affiliation{Department of Physics, University of Wisconsin--Madison, 1150 University Avenue, Madison, WI 53706, USA}

\date{\today}

\begin{abstract}
A novel plasma equilibrium in the high-$\beta$, Hall regime that produces centrally-peaked, high Mach number Couette flow is described. Flow is driven using a weak, uniform magnetic field and large, cross field currents. Large magnetic field amplification (factor 20) due to the Hall effect is observed when electrons are flowing radially inward, and near perfect field expulsion is observed when the flow is reversed. A dynamic equilibrium is reached between the amplified (removed) field and extended density gradients.  
\end{abstract}

\maketitle

Fluid flow between two concentric cylinders, Taylor-Couette flow~\footnote{For brevity we will shorten this to ``Couette" for the rest of the letter.}, has been a cornerstone of fluid mechanics for more than 300 years~\cite{Donnelly1991}. Starting with Newton, this simple geometry has served as theoretical and experimental platform for hydrodynamics. Couette flow served as basis for the design of the earliest viscometers, where the name Couette comes from~\cite{Mallock, Couette}. It has been a major tool in modern studies of fluid turbulence, particularly the pioneering work of Taylor~\cite{Taylor}. Extending beyond conventional fluids, Couette flow has been used to characterize more complex media such as visco-elastic polymers \cite{Boldyrev,Graham} and magneto-fluids such as liquid metals, where the flowing fluid is subject to electromagnetic forces in addition to pressure and viscosity. Chandrasekhar and Velikhov simultaneously described the stability of magneto-fluid Couette flow in the presence of weak magnetic fields~\cite{Chandrasekhar, Velikhov1959}. Most recently, Couette flow of unmagnetized plasma has been realized in the lab and provided for measurements of plasma viscosity~\cite{cami_prl}.

Due to the similarity to Keplerian flow ($V_{\phi}\propto r^{-1/2}$), Couette flow has been proposed as model system for laboratory astrophysics. There have been numerous attempts to experimentally study the magnetorotational instability (MRI) in liquid metal experiments \cite{Ji2001, Noguchi2002, Lathrop2004, Gissinger2011, Mark2010, Stefani2006, Liu2006}, but these efforts have often been met with complications caused by parasitic modes from the boundaries \cite{Szklarski2007, Roach2012, Spence2012}. Plasma Couette experiments open up MRI research to kinetic physics, Hall effects and mixed charged-neutral systems, highlighting issues that are import in hot, dense disks \cite{Quataert2002, Kunz2016, Kunz2019} as well as partially-magnetized protostellar systems \cite{wardle1999, WardleNg1999, Balbus2001, Ebrahimi2011, Kunz2013, Lesur2014}. The Hall effect has astrophysical applications beyond the MRI; such as Hall dynamos~\cite{Mininni_2002, Mininni_2003}, turbulent reconnection in the magnetosphere~\cite{Phan2018}, and magnetic field evolution in the crusts of neutron stars~\cite{Cumming2004, Goldreich1992, Rheinhardt2002}. In principle, any dynamics where the ions are decoupled from the magnetic field need to be treated with the Hall term in Ohm's law. For an experiment this occurs when the ion inertial length is comparable to the system size, while in extremely large astrophysical systems, Hall dynamics are mostly restricted to smaller scale phenomena.

\begin{figure}
    \centering
    \includegraphics{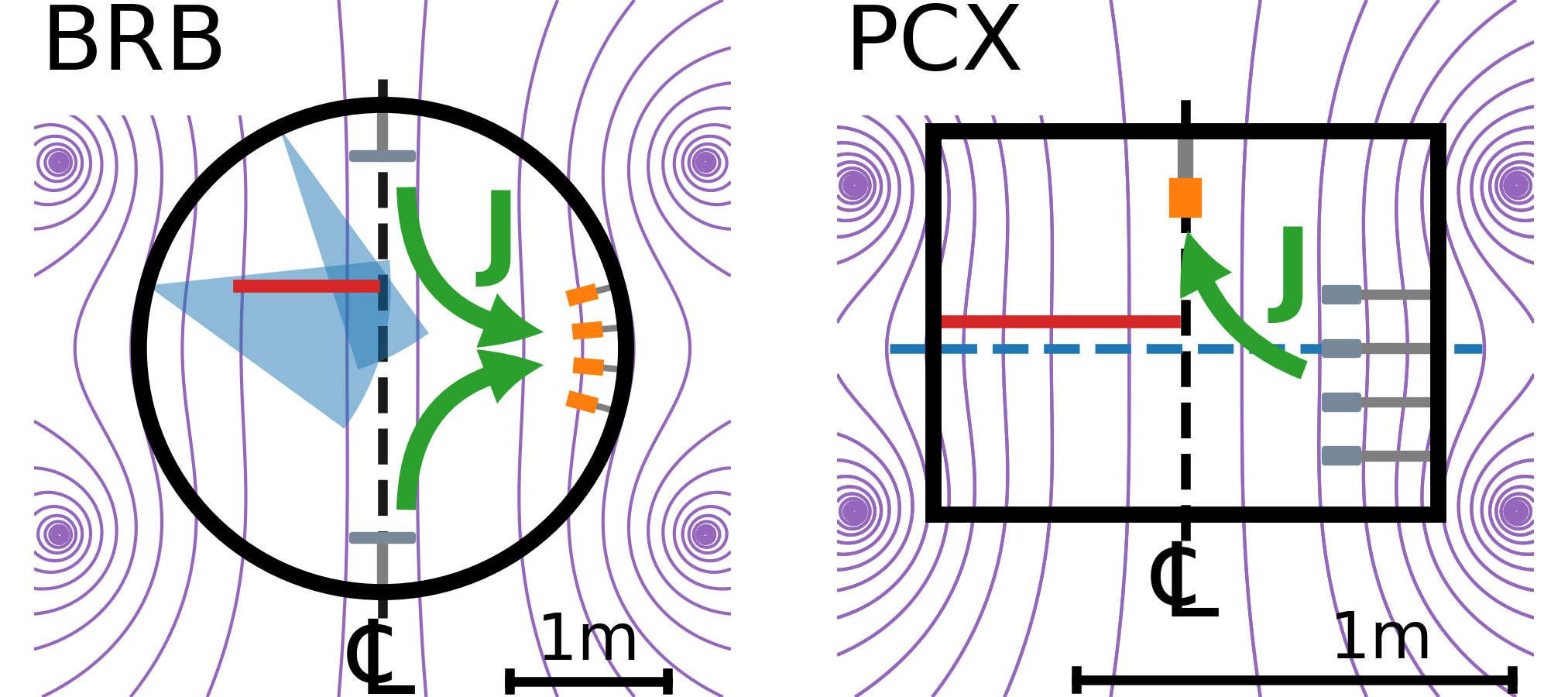}
    \caption{Volumetric Flow Drive implemented on (Left) the BRB with outward current and (Right) on PCX with inward flowing current. The schematics show the electrode positions and probe scanning locations (blue wedges and red line) for the BRB and the blue dashed line shows the location of the Fabry-Perot optical diagnostic that measures chord integrated ion temperature and flow on PCX. Both devices are axisymmetric about the rotation axis. In both configurations current is sourced by LaB$_6$ cathodes (orange) and collected by cold molybdenum anodes (gray). A weak ($<10$~G) externally applied magnetic field (purple) assures the plasma has an initial $\beta > 10$ and that the ions are weakly magnetized.}
    \label{fig:schematic}
\end{figure}

Plasma Couette experiments have been mostly performed in the unmagnetized limit, where no external field is applied, mostly to ensure that there is enough viscous transport from the driving boundaries to spin up the entire volume~\cite{cami_prl, pcx_PoP}. Additionally, to our knowledge, there has been no experimental evidence supporting the necessary inclusion of two-fluid effects in astrophysical plasmas outside of magnetic reconnection contexts. In this letter, we present a novel dynamic equilibrium that produces plasma Couette flow in the weakly-magnetized, high-$\beta$~\footnote{$\beta$ is the ratio of plasma to magnetic pressure commonly used amongst plasma physicists.}, Hall regime. In our system, flow is driven via a body force applied across the entire volume that relies on a weak applied magnetic field and a cross-field current. This so-called volumetric flow drive (VFD) has been considered for use in MRI experiments in liquid metals \cite{Khalzov2006, Velikhov2006, Khalzov2010} and proof-of-principle experiments have suggested that flow drive is possible in plasmas \cite{dave_flow}. Here, however, we show that certain configurations of VFD in Hall plasmas result in a massive amplification of the initial field by the Hall effect and a hollowing out of the density profile with centrally peaked flows--radically altering the equilibrium state expected from the MHD model. After presenting a description of the experimental setup, we show equilibrium measurements of plasma VFD in two configurations: outwardly and inwardly directed current. The outwardly directed current case shows strong field amplification, hollow density profiles and Couette flow, while the opposite case shows strong field expulsion and solid-body flow profiles. We then compare these measurements to extended MHD simulations in order to develop a simple two-fluid model of this equilibrium. 

The experiments presented here were carried out in two very similar devices, the Big Red Ball (BRB) and the Plasma Couette Experiment (PCX), both operated at the Wisconsin Plasma Physics Laboratory (WiPPL) \cite{wippl_JPP, mpdx_PoP, cami_prl}. Plasma creation and flow drive are achieved by injecting current from hot, emissive lanthanum hexaboride cathodes (LaB$_6$) across a weak, externally applied magnetic field \cite{DaveThesis, DavePoPFlow}. Due to the multi-cusp confinement scheme for both devices, high-$\beta$ can be achieved with ion inertial lengths ($d_i = c/\Omega_{pi}$) on the order of 1~m, which places these devices firmly in the Hall regime. Argon plasmas are produced by injecting 30-300~A of current from the LaB$_6$ cathodes with a constant neutral fill of approximately $10^{-5}$~torr. These discharges reach densities on the order of 10$^{17}$-10$^{18}$~m$^{-3}$, electron temperatures of 3-5~eV, and ion temperatures of 0.5-1.5~eV. With the weak applied fields in the range of 0.3-10~G, the electrons are able to execute many gyro-orbits between collisions, while the ion gyroradius is on the order of the device size.

Figure~\ref{fig:schematic} shows a diagram of the flow scheme for both devices. The BRB is a spherical device, roughly 3~m in diameter, while PCX is cylindrical and roughly 1~m diameter, 1~m tall. For the BRB, current is driven radially outward using a set of 6 cathodes and two large ring anodes placed near the poles. In PCX, the current is driven radially inward with a single cathode on axis and 4 anodes located near the edge. In terms of the flow rotation vector, $\mathbf{\Omega}$, BRB operates with $\mathbf{B}\nparallel\mathbf{\Omega}$ (antiparallel), while PCX has $\mathbf{B}\parallel\mathbf{\Omega}$. This is true regardless of the direction of the applied magnetic field, since the rotation is set by the $\mathbf{J}\times\mathbf{B}$ torque. By having both orientations, we are able to compare large qualitative differences in the resulting equilibria.

The magnetic field is measured by a calibrated 15-position, 3-axis Hall probe array with a resolution of approximately 0.1~G \cite{Ethan_Thesis}. Density, electron temperature, and flow are measured by a single position combination Mach and Langmuir probe using standard analysis techniques. Both Hall and electrostatic probes are spatially scanned over the areas indicated in Fig.\ref{fig:schematic} over the course of many shots, with fixed electrostatic probes used to determine shot-to-shot reproducibility. In addition to probes, PCX is equipped with a unique, high-resolution Fabry-Perot spectrometer, which is able to measure chord integrated ion temperature and flow to better than 0.1~eV and 50~m/s precision \cite{jason_rsi}. 

\begin{figure}
    \centering
    \includegraphics[width=8.6cm]{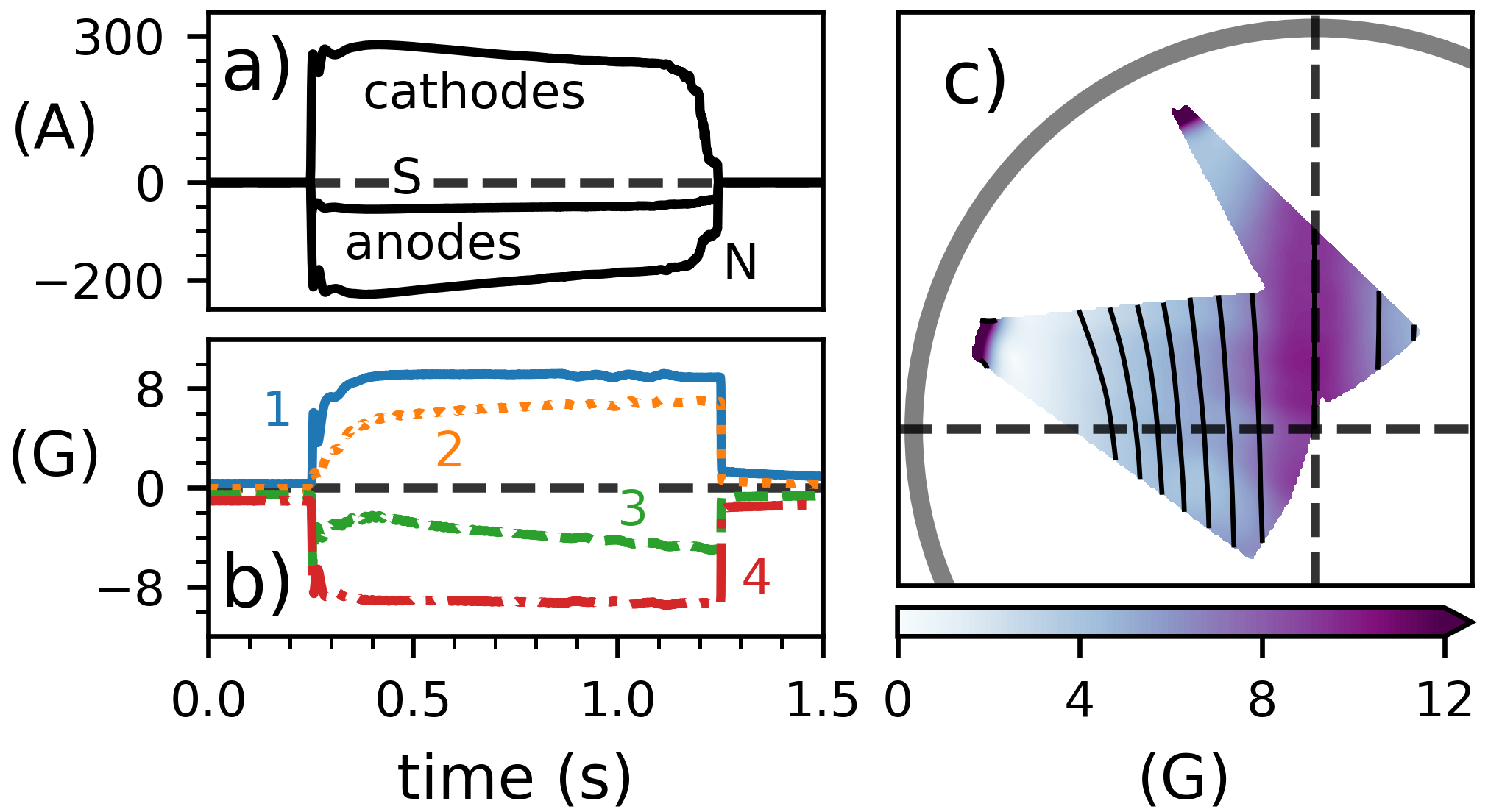}
    \caption{Data from the BRB. a) time traces of the electrode currents b) time traces of $B_{z}$ taken near the axis of the machine for four different initial field cases (0.4~G, 0.1~G, -0.6~G, -1.6~G). c) Poloidal map of the magnetic field strength with lines of flux corresponding to case 1 from b). The initial field is amplified by a factor of 20 on axis.}
    \label{fig:brb_data}
\end{figure}

Figure~\ref{fig:brb_data} shows BRB data where outward directed current drives large magnetic field amplification. A bias is applied at $t=0.25$~s between the LaB$_6$ cathodes and the polar anodes that generates a second-long steady plasma. Immediately after the plasma is created, massive field amplification is observed. Figure.~\ref{fig:brb_data}(b) shows the amplification of four different initial magnetic field cases. In each of these cases, the initial vacuum field is amplified by at least a factor of 10 with roughly 300~A of total injected current, regardless of field direction or magnitude. Focusing on case 1, corresponding to an initial field of $\sim0.4$~G, Fig.~\ref{fig:brb_data}(c) shows the poloidal map of the magnetic field strength and associated field lines indicating the amplification is concentrated on axis and is uniform in the axial direction. A weak, $<1$~G, toroidal magnetic field was measured as well, which is consistent with the poloidal currents being driven in the system.

\begin{figure}
    \centering
    \includegraphics[width=8.6cm]{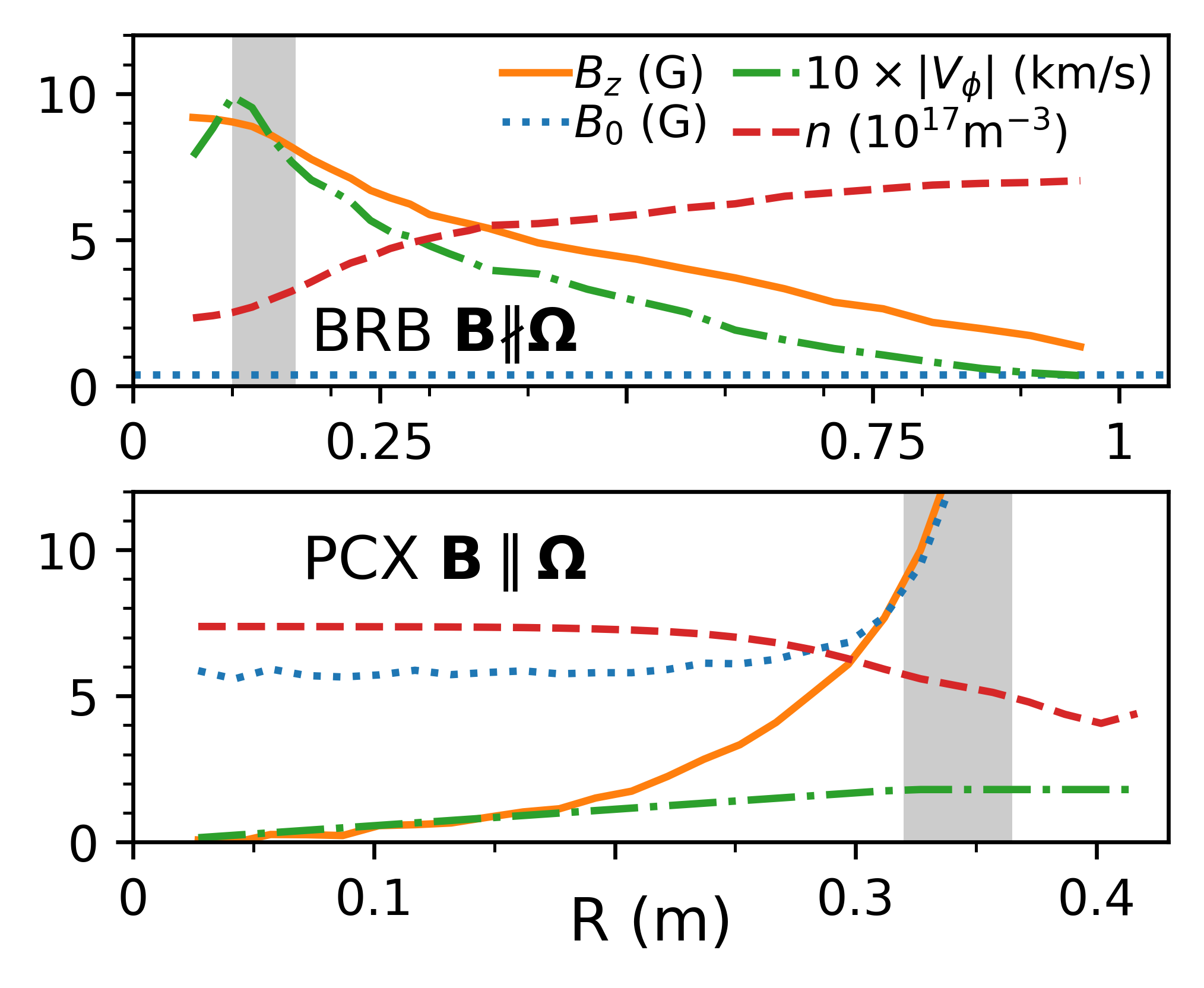}
    \caption{Linear profiles of the magnetic field, toroidal flow and density during the plasma discharge for BRB and PCX. The shaded vertical bars indicate the radii of the anodes. The velocity profile for PCX was measured using the Fabry-P\'erot spectrometer. On the BRB, these profiles are measured along the cylindrical radius indicated by the red line in Fig.~\ref{fig:schematic}, approximately 40~cm from the equator. On PCX, the Fabry-P\'erot flow measurement and density profile are made at the mid-plane and the magnetic field profiles are measured slightly above the mid-plane (indicated by the red line in Fig.~\ref{fig:schematic}).}
    \label{fig:lin_prof}
\end{figure}

A hollow density profile accompanies the large field amplification. Figure~\ref{fig:lin_prof}(top) shows radial profiles of the magnetic field, flow, and density from the BRB during a plasma discharge. The density near the axis of rotation is reduced by more than a factor of 2 from the bulk. In most BRB discharges using the LaB$_6$ cathodes, the density is uniform throughout the bulk volume, with only a short gradient near the magnetic cusp at the edge \cite{mpdx_PoP}. For cases with larger field amplification (1 and 4 in Fig.~\ref{fig:brb_data}b), the density gradient is steeper than the lower field cases. In all cases, the equilibrium $\beta$ near the center was still $>0.5$ despite the hollow density profile and amplified field.

Strong, centrally peaked, Couette flow is also driven in this equilibrium. The flow profile peaks near the radial anode location, approximately 10~cm from the axis. In the no dissipation limit, this centrally peaked flow profile has enough shear to meet the Rayleigh circulation criteria for hydrodynamic instability as well as the ideal MRI requirement\cite{Rayleigh1879, balbus1991}, however no characteristics of instability are observed. 

When the current direction is reversed using PCX, the magnetic field is removed from the plasma. Figure~\ref{fig:lin_prof}~(lower) shows linear profiles from the reversed current case on PCX. The roughly 7~G initial magnetic field is completely removed from the central region of the plasma. Along with the field removal, an elongated density gradient is seen that extends from the plasma edge well into the bulk volume. This gradient is significantly longer than the typical one seen from the multi-cusp confinement ($\sim$10~cm)~\cite{Cooper2016_cusp}.

Flow is also observed on PCX, but instead of being centrally peaked, it closely follows a solid body profile as measured by the Fabry-P\'erot spectrometer~\cite{jason_rsi}. Previous experiments on PCX showed solid-body-like flow was created by locally stirring the plasma near the outer edge and relying on strong unmagnetized viscosity of the interior to transport momentum inward~\cite{cami_prl}. The observations here are similar to these flow profiles, suggesting that the torque is local and near the outer boundary where the magnetic field is strongest. The flow is also much smaller than on BRB, most likely due to the stronger cusp field ($>20$~G) in the drive region near the edge. 

\begin{figure}
    \centering
    \includegraphics[width=8.6cm]{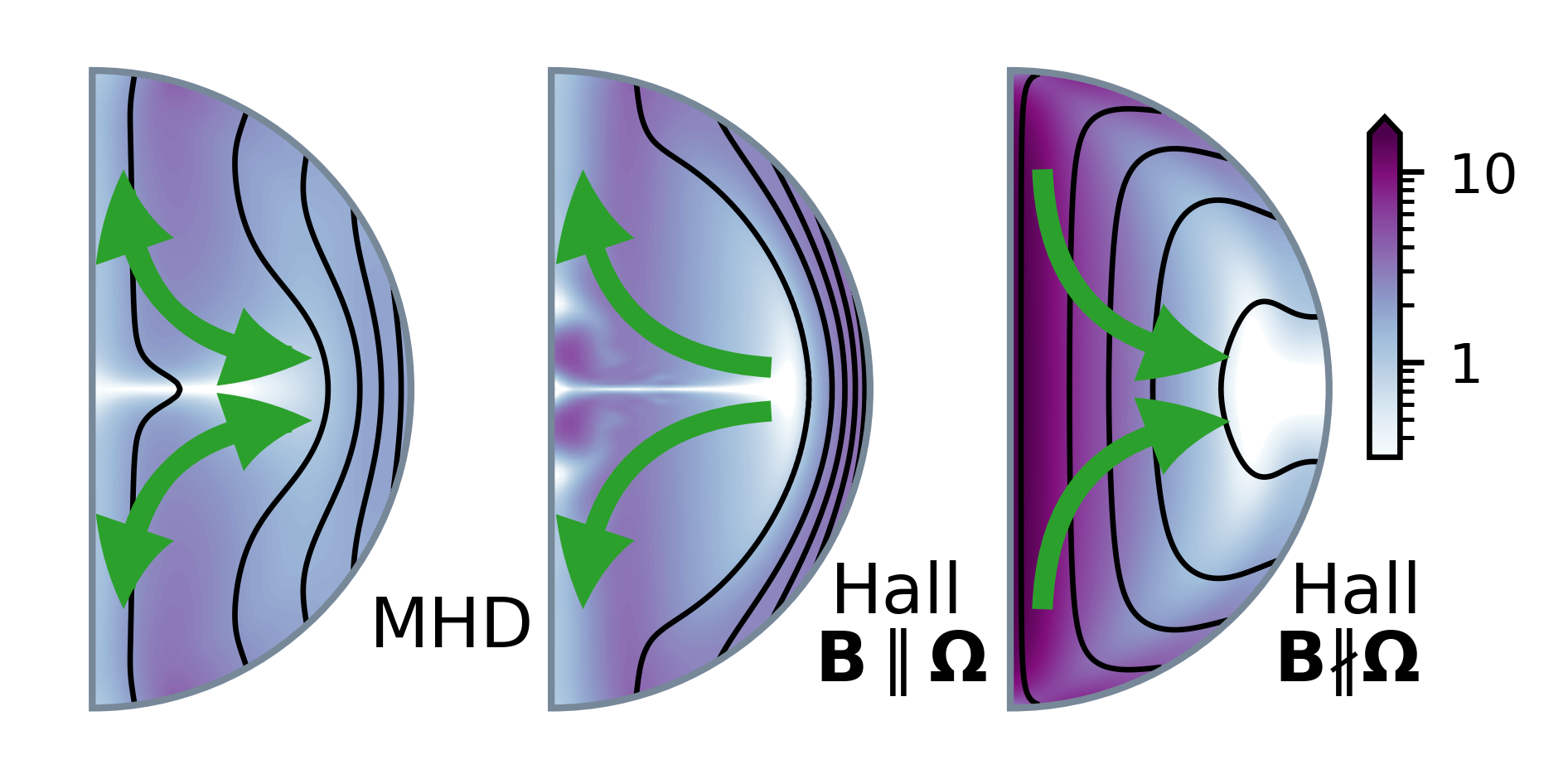}
    \caption{$|{\bf B}|/B_{0}$ and flux lines for NIMROD simulations. In all cases the initial field is uniform and has a strength of 0.5~G. The contour is plotted on a log scale. Left: Case without Hall terms in Ohm's law. Middle: Case with Hall terms included in Ohm's law and the current directed inward, which is the same direction as the PCX experiments. Right: Hall case with the current directed outward, which is the same as the experimental BRB case.}
    \label{fig:nimrod}
\end{figure}

In conjunction with the observations on BRB and PCX, simulations using the NIMROD extended MHD solver \cite{nimrod} help to clarify the role of the Hall effect. NIMROD performs semi-implicit time stepping on a poloidal finite element grid, with a spectral representation for the toroidal coordinate. Current injection is modeled by setting the toroidal magnetic field along the boundary, which is equivalent to setting a radial current \cite{Ethan_Nature}. NIMROD operates with fully-conductive, flux conserving boundaries, so the injected current is not a source of magnetic flux but still serves as an excellent parallel to the experiment. In these simulations, typical BRB parameters for a helium plasma ($n\sim6\times10^{17}$~m$^{-3}$, $T_{e}=8$~eV, $T_{i}=0.5$~eV) are used and the total injected current is 400~A. These inputs clearly highlight the qualitative effects seen in the experiment, but do not allow for a direct quantitative comparison.

A series of simulations are preformed that selectively include two-fluid terms from the generalized Ohm's law. For this discussion, Ohm's law with only fluid induction and resistivity is labeled ``MHD", while cases where current induction and electron pressure are included are labeled as ``Hall''. Figure~\ref{fig:nimrod} shows the magnetic field relative to the applied field and flux lines for four separate cases. In the MHD case, Ekman circulation develops and drives a radial outward flow which drags the field lines in either current direction. When the Hall term is included, the field is coupled to the electron fluid, where the direction of radial flow is determined by the applied current. The simulations confirm that volumetric flow drive only amplifies field with the extended Ohm's law terms included and outwardly directed radial current, corresponding to $\mathbf{B}\nparallel\mathbf{\Omega}$. In Hall runs with the opposite current direction, the field is mostly removed from the bulk of the plasma volume, matching the observations made on PCX. 

The Hall effect mechanism responsible for the field amplification or removal can be easily seen by considering the extended Ohm's law,
\begin{equation}
    {\bf E} - {\bf V}\times{\bf B} = \eta {\bf J} + \frac{1}{ne}\left({\bf J}\times{\bf B} - \nabla P_{e}\right)
    \label{eq:ohms}
\end{equation}
where $\eta$ is the plasma resistivity and $P_{e}\equiv nkT_{e}$ is the electron pressure. In the high-$\beta$ Hall limit, the ${\bf J}\times{\bf B}$ and electron pressure terms dominate in balancing the applied $E_r$. By considering the toroidal component of Ohm's law and setting the non-equilibrium inductive electric field to zero, a relationship between the radial and toroidal currents is found,
\begin{equation}
    E_{\phi}=0=\eta J_{\phi} - \frac{1}{ne}J_{r}B_{z}\hspace{5mm}\rightarrow\hspace{5mm}J_{\phi}=\frac{\Omega_{ce}}{\nu_{e}} J_{r}
    \label{eq:jphi}
\end{equation}
where the Spitzer form of resistivity has been used to relate the resistivity to the electron collision frequency, $\nu_{e}$. In cases where the electrons are well magnetized relative to their collisions, a large toroidal current can be formed from cross-field current. Applying Ampere's law, when $J_{r}>0$ (as on the BRB) the induced $J_{\phi}$ will always act to enforce the magnetic field, while when $J_{r}<0$ (as on PCX), the toroidal current will act against the existing magnetic field. 

An ordering for the terms in Eq.~\ref{eq:ohms} for parameters in either device indicates that the flow induction and resistivity terms are negligible compared to the current induction and electron pressure. The resistivity term is kept to arrive at the expression in Eq.~\ref{eq:jphi} because some collisions are required for cross-field current. However, for both species, collisions do not play a large role in the radial force balance and have been neglected in this simple model. 

While the electrons are well magnetized and drifting to create strong toroidal currents, the ions are mostly unmagnetized and ballistic. However, the ions still play an important role in managing the toroidal current since their force balance sets the density profile. In the absence of the Lorentz force, a radial electric field sets up to balance the centrifugal force from ion flow and the ion pressure gradient, 
\begin{equation}
    -nm_{i}\frac{V_{\phi}^{2}}{r} = neE_{r} - \frac{\partial P_{i}}{\partial r}
    \label{eq:ion}
\end{equation}
where $P_{i}\equiv kT_{i}n$ is the ion pressure. For these plasmas the terms on the RHS of equation \ref{eq:ion} dominate and so the ion pressure gradient is largely balanced by the radial electric field and the ions can be thought of a Boltzmann-like in this equilibrium. When the electric field is outwardly directed (like on BRB), the density profile is hollow, while an inwardly directed electric field (like on PCX) causes the density to peak on axis. This electric field couples the ions and electrons, completing the equilibrium model.

Attributing plasma current entirely to the electrons due to their high mobility and using the electric field from Eq.~\ref{eq:ion}, leads to the standard plasma equilibrium condition in the radial direction,
\begin{equation}
    J_{\phi}B_{z} = \frac{\partial}{\partial r}(P_{i}+P_{e}) - nm_{i}\frac{V_{\phi}^{2}}{r}
    \label{eq:equil}
\end{equation}
where the last term is a small correction arising from the ion flow. This standard equilibrium coupled with the Hall mechanism in Eq.~\ref{eq:jphi} shows that the generated current necessarily causes extended density gradients and that the direction of the gradient is dependent on the injected current direction (electric field in the ion force balance). 

\begin{figure}
    \centering
    \includegraphics[width=8.6cm]{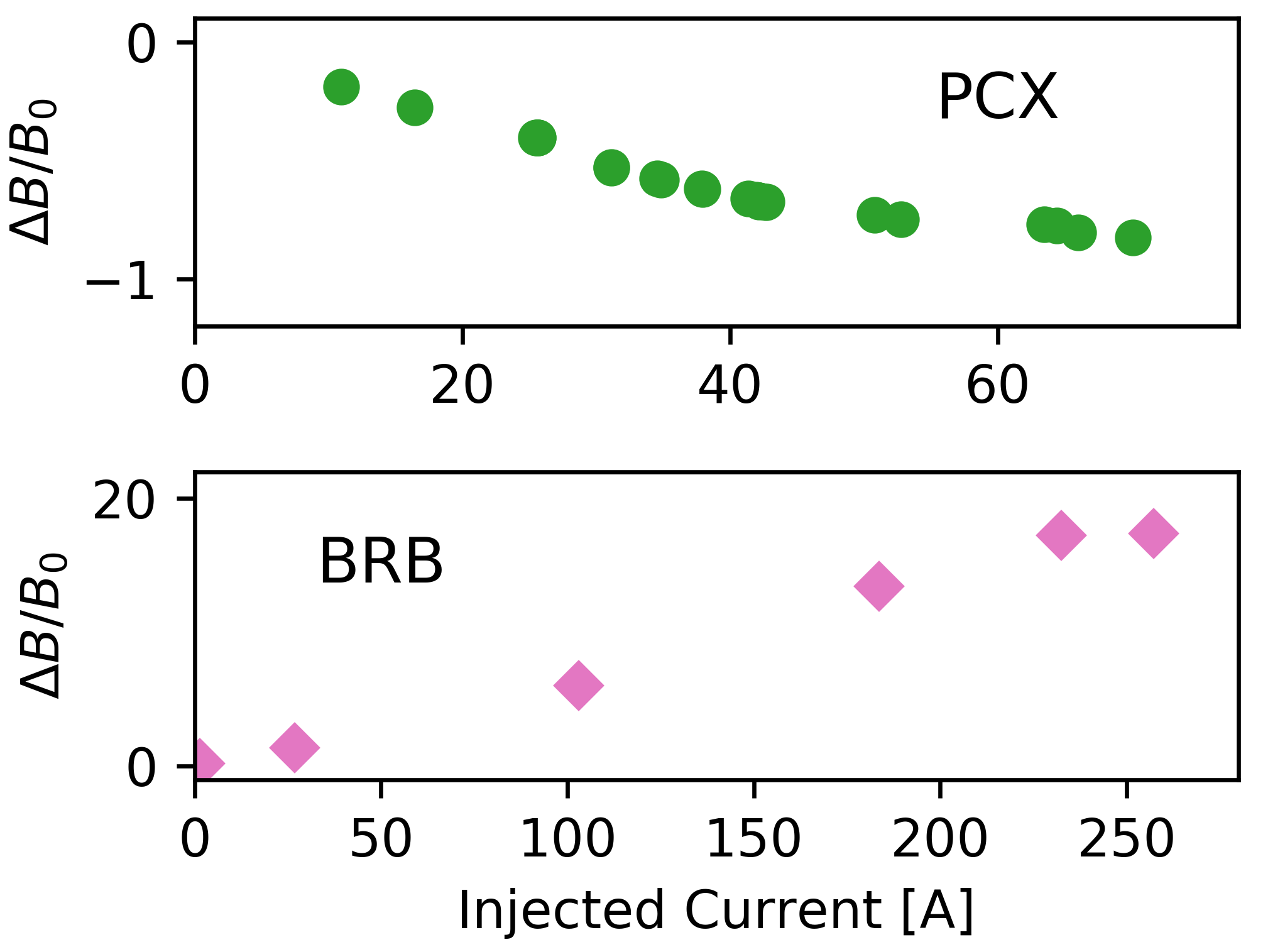}
    \caption{Scan of injected current versus normalized change in magnetic field for the two experiments. Top: The PCX case shows strong diamagnetic field removal, approaching total removal at approximately 80~A of injected current. For the BRB case, the field is amplified by nearly a factor of 20 at nearly 300~A of injected current.}
    \label{fig:current_scale}
\end{figure}

The equilibrium described here takes advantage of the well-confined plasmas in multicusp devices, where an ambipolar field in the small cusp region keeps the ions from leaving the plasma. In previous work with similar flux expulsion experiments~\cite{Stenzel2000}, an ad hoc radial electric field is used to complete the electron force balance. Here however, this electric field is well described by Eq.~\ref{eq:ion}, leading to the standard MHD force balance in Eq.~\ref{eq:equil}. The coupling of the electrons and ions via the electric field is an essential feature of this Hall framework. 

In both configurations the total amount of poloidal magnetic flux was not conserved during the progression of the experiments. This apparent creation or annihilation of flux is the result of the Hall effect's conversion of the flux carried by the injected current into poloidal flux in the plasma. On BRB, the magnetic field is amplified by nearly a factor of 20 at the maximum injected current of nearly 300~A (see bottom of Fig.~\ref{fig:current_scale}). On PCX, the injected current was scanned, showing a positive correlation between the amount of removed flux and the total injected current (see top of Fig.~\ref{fig:current_scale}). In both cases, as more current is injected and converted into poloidal flux by the Hall mechanism in Eq.~\ref{eq:jphi}, more field is added or removed from the plasma. 

In summary, the present study demonstrates a new type of plasma flow drive, similar to Couette flow, that uses cross-field currents to drive cylindrically symmetric plasmas with sheared flows. We have shown conclusively that {\sl a priori} unimportant and weak ($\beta\gg1$) magnetic fields can, in fact, greatly influence the large-scale equilibrium via the Hall effect. Similar conclusions have been noted in the case of magnetic reconnection where Hall effects can control large scale dynamics by influencing the scale where the magnetic field vanishes~\cite{Birn2001,JoePRL, MRX_Hall}. To our knowledge, our results represent the first experimental evidence of large qualitative Hall effects in a laboratory astrophysics context outside of magnetic reconnection.

\begin{acknowledgments}
This work was funded in part by the NSF under grant No. 1518115 and by the D.O.E. under grant No. DE-SC0018266.
\end{acknowledgments}

\providecommand{\noopsort}[1]{}\providecommand{\singleletter}[1]{#1}%


\begin{thebibliography}{54}%
\makeatletter
\providecommand \@ifxundefined [1]{%
 \@ifx{#1\undefined}
}%
\providecommand \@ifnum [1]{%
 \ifnum #1\expandafter \@firstoftwo
 \else \expandafter \@secondoftwo
 \fi
}%
\providecommand \@ifx [1]{%
 \ifx #1\expandafter \@firstoftwo
 \else \expandafter \@secondoftwo
 \fi
}%
\providecommand \natexlab [1]{#1}%
\providecommand \enquote  [1]{``#1''}%
\providecommand \bibnamefont  [1]{#1}%
\providecommand \bibfnamefont [1]{#1}%
\providecommand \citenamefont [1]{#1}%
\providecommand \href@noop [0]{\@secondoftwo}%
\providecommand \href [0]{\begingroup \@sanitize@url \@href}%
\providecommand \@href[1]{\@@startlink{#1}\@@href}%
\providecommand \@@href[1]{\endgroup#1\@@endlink}%
\providecommand \@sanitize@url [0]{\catcode `\\12\catcode `\$12\catcode
  `\&12\catcode `\#12\catcode `\^12\catcode `\_12\catcode `\%12\relax}%
\providecommand \@@startlink[1]{}%
\providecommand \@@endlink[0]{}%
\providecommand \url  [0]{\begingroup\@sanitize@url \@url }%
\providecommand \@url [1]{\endgroup\@href {#1}{\urlprefix }}%
\providecommand \urlprefix  [0]{URL }%
\providecommand \Eprint [0]{\href }%
\providecommand \doibase [0]{https://doi.org/}%
\providecommand \selectlanguage [0]{\@gobble}%
\providecommand \bibinfo  [0]{\@secondoftwo}%
\providecommand \bibfield  [0]{\@secondoftwo}%
\providecommand \translation [1]{[#1]}%
\providecommand \BibitemOpen [0]{}%
\providecommand \bibitemStop [0]{}%
\providecommand \bibitemNoStop [0]{.\EOS\space}%
\providecommand \EOS [0]{\spacefactor3000\relax}%
\providecommand \BibitemShut  [1]{\csname bibitem#1\endcsname}%
\let\auto@bib@innerbib\@empty
\bibitem [{\citenamefont {Donnelly}(1991)}]{Donnelly1991}%
  \BibitemOpen
  \bibfield  {author} {\bibinfo {author} {\bibfnamefont {R.~J.}\ \bibnamefont
  {Donnelly}},\ }\bibfield  {title} {\bibinfo {title} {Taylor--couette flow:
  the early days},\ }\href@noop {} {\bibfield  {journal} {\bibinfo  {journal}
  {Phys. Today}\ }\textbf {\bibinfo {volume} {44}},\ \bibinfo {pages} {32}
  (\bibinfo {year} {1991})}\BibitemShut {NoStop}%
\bibitem [{\citenamefont {Mallock}(1889)}]{Mallock}%
  \BibitemOpen
  \bibfield  {author} {\bibinfo {author} {\bibfnamefont {A.}~\bibnamefont
  {Mallock}},\ }\bibfield  {title} {\bibinfo {title} {Iv. determination of the
  viscosity of water},\ }\href@noop {} {\bibfield  {journal} {\bibinfo
  {journal} {Proceedings of the Royal Society of London}\ }\textbf {\bibinfo
  {volume} {45}},\ \bibinfo {pages} {126} (\bibinfo {year} {1889})}\BibitemShut
  {NoStop}%
\bibitem [{\citenamefont {{Couette, M.}}(1890)}]{Couette}%
  \BibitemOpen
  \bibfield  {author} {\bibinfo {author} {\bibnamefont {{Couette, M.}}},\
  }\bibfield  {title} {\bibinfo {title} {Distinction de deux r\'egimes dans le
  mouvement des fluides},\ }\href
  {https://doi.org/10.1051/jphystap:018900090041401} {\bibfield  {journal}
  {\bibinfo  {journal} {J. Phys. Theor. Appl.}\ }\textbf {\bibinfo {volume}
  {9}},\ \bibinfo {pages} {414} (\bibinfo {year} {1890})}\BibitemShut {NoStop}%
\bibitem [{\citenamefont {Taylor}(1923)}]{Taylor}%
  \BibitemOpen
  \bibfield  {author} {\bibinfo {author} {\bibfnamefont {G.~I.}\ \bibnamefont
  {Taylor}},\ }\bibfield  {title} {\bibinfo {title} {Stability of a viscous
  liquid contained between two rotating cylinders},\ }\href
  {http://www.jstor.org/stable/91148} {\bibfield  {journal} {\bibinfo
  {journal} {Philosophical Transactions of the Royal Society of London. Series
  A, Containing Papers of a Mathematical or Physical Character}\ }\textbf
  {\bibinfo {volume} {223}},\ \bibinfo {pages} {289} (\bibinfo {year}
  {1923})}\BibitemShut {NoStop}%
\bibitem [{\citenamefont {Boldyrev}\ \emph {et~al.}(2009)\citenamefont
  {Boldyrev}, \citenamefont {Huynh},\ and\ \citenamefont {Pariev}}]{Boldyrev}%
  \BibitemOpen
  \bibfield  {author} {\bibinfo {author} {\bibfnamefont {S.}~\bibnamefont
  {Boldyrev}}, \bibinfo {author} {\bibfnamefont {D.}~\bibnamefont {Huynh}},\
  and\ \bibinfo {author} {\bibfnamefont {V.}~\bibnamefont {Pariev}},\
  }\bibfield  {title} {\bibinfo {title} {Analog of astrophysical
  magnetorotational instability in a couette-taylor flow of polymer fluids},\
  }\href {https://doi.org/10.1103/PhysRevE.80.066310} {\bibfield  {journal}
  {\bibinfo  {journal} {Phys. Rev. E}\ }\textbf {\bibinfo {volume} {80}},\
  \bibinfo {pages} {066310} (\bibinfo {year} {2009})}\BibitemShut {NoStop}%
\bibitem [{\citenamefont {GRAHAM}(1998)}]{Graham}%
  \BibitemOpen
  \bibfield  {author} {\bibinfo {author} {\bibfnamefont {M.~D.}\ \bibnamefont
  {GRAHAM}},\ }\bibfield  {title} {\bibinfo {title} {Effect of axial flow on
  viscoelastic taylor?couette instability},\ }\href
  {https://doi.org/10.1017/S0022112098008611} {\bibfield  {journal} {\bibinfo
  {journal} {Journal of Fluid Mechanics}\ }\textbf {\bibinfo {volume} {360}},\
  \bibinfo {pages} {341?374} (\bibinfo {year} {1998})}\BibitemShut {NoStop}%
\bibitem [{\citenamefont {Chandrasekhar}(1960)}]{Chandrasekhar}%
  \BibitemOpen
  \bibfield  {author} {\bibinfo {author} {\bibfnamefont {S.}~\bibnamefont
  {Chandrasekhar}},\ }\bibfield  {title} {\bibinfo {title} {The stability of
  non-dissipative couette flow in hydromagnetics},\ }\href@noop {} {\bibfield
  {journal} {\bibinfo  {journal} {Proceedings of the National Academy of
  Sciences of the United States of America}\ }\textbf {\bibinfo {volume}
  {46}},\ \bibinfo {pages} {253} (\bibinfo {year} {1960})}\BibitemShut
  {NoStop}%
\bibitem [{\citenamefont {Velikhov}(1959)}]{Velikhov1959}%
  \BibitemOpen
  \bibfield  {author} {\bibinfo {author} {\bibfnamefont {E.}~\bibnamefont
  {Velikhov}},\ }\bibfield  {title} {\bibinfo {title} {Stability of an ideally
  conducting liquid flowing between cylinders rotating in a magnetic field},\
  }\href@noop {} {\bibfield  {journal} {\bibinfo  {journal} {Sov. Phys. JETP}\
  }\textbf {\bibinfo {volume} {36}},\ \bibinfo {pages} {995} (\bibinfo {year}
  {1959})}\BibitemShut {NoStop}%
\bibitem [{\citenamefont {Collins}\ \emph {et~al.}(2012)\citenamefont
  {Collins}, \citenamefont {Katz}, \citenamefont {Wallace}, \citenamefont
  {Jara-Almonte}, \citenamefont {Reese}, \citenamefont {Zweibel},\ and\
  \citenamefont {Forest}}]{cami_prl}%
  \BibitemOpen
  \bibfield  {author} {\bibinfo {author} {\bibfnamefont {C.}~\bibnamefont
  {Collins}}, \bibinfo {author} {\bibfnamefont {N.}~\bibnamefont {Katz}},
  \bibinfo {author} {\bibfnamefont {J.}~\bibnamefont {Wallace}}, \bibinfo
  {author} {\bibfnamefont {J.}~\bibnamefont {Jara-Almonte}}, \bibinfo {author}
  {\bibfnamefont {I.}~\bibnamefont {Reese}}, \bibinfo {author} {\bibfnamefont
  {E.}~\bibnamefont {Zweibel}},\ and\ \bibinfo {author} {\bibfnamefont {C.~B.}\
  \bibnamefont {Forest}},\ }\bibfield  {title} {\bibinfo {title} {Stirring
  unmagnetized plasma},\ }\href
  {https://doi.org/10.1103/PhysRevLett.108.115001} {\bibfield  {journal}
  {\bibinfo  {journal} {Phys. Rev. Lett.}\ }\textbf {\bibinfo {volume} {108}},\
  \bibinfo {pages} {115001} (\bibinfo {year} {2012})}\BibitemShut {NoStop}%
\bibitem [{\citenamefont {Ji}\ \emph {et~al.}(2001)\citenamefont {Ji},
  \citenamefont {Goodman},\ and\ \citenamefont {Kageyama}}]{Ji2001}%
  \BibitemOpen
  \bibfield  {author} {\bibinfo {author} {\bibfnamefont {H.}~\bibnamefont
  {Ji}}, \bibinfo {author} {\bibfnamefont {J.}~\bibnamefont {Goodman}},\ and\
  \bibinfo {author} {\bibfnamefont {A.}~\bibnamefont {Kageyama}},\ }\bibfield
  {title} {\bibinfo {title} {{Magnetorotational instability in a rotating
  liquid metal annulus}},\ }\href
  {https://doi.org/10.1046/j.1365-8711.2001.04647.x} {\bibfield  {journal}
  {\bibinfo  {journal} {Monthly Notices of the Royal Astronomical Society}\
  }\textbf {\bibinfo {volume} {325}},\ \bibinfo {pages} {L1} (\bibinfo {year}
  {2001})}\BibitemShut {NoStop}%
\bibitem [{\citenamefont {Noguchi}\ \emph {et~al.}(2002)\citenamefont
  {Noguchi}, \citenamefont {Pariev}, \citenamefont {Colgate}, \citenamefont
  {Beckley},\ and\ \citenamefont {Nordhaus}}]{Noguchi2002}%
  \BibitemOpen
  \bibfield  {author} {\bibinfo {author} {\bibfnamefont {K.}~\bibnamefont
  {Noguchi}}, \bibinfo {author} {\bibfnamefont {V.~I.}\ \bibnamefont {Pariev}},
  \bibinfo {author} {\bibfnamefont {S.~A.}\ \bibnamefont {Colgate}}, \bibinfo
  {author} {\bibfnamefont {H.~F.}\ \bibnamefont {Beckley}},\ and\ \bibinfo
  {author} {\bibfnamefont {J.}~\bibnamefont {Nordhaus}},\ }\bibfield  {title}
  {\bibinfo {title} {Magnetorotational instability in liquid metal couette
  flow},\ }\href {https://doi.org/10.1086/341502} {\bibfield  {journal}
  {\bibinfo  {journal} {The Astrophysical Journal}\ }\textbf {\bibinfo {volume}
  {575}},\ \bibinfo {pages} {1151} (\bibinfo {year} {2002})}\BibitemShut
  {NoStop}%
\bibitem [{\citenamefont {Sisan}\ \emph {et~al.}(2004)\citenamefont {Sisan},
  \citenamefont {Mujica}, \citenamefont {Tillotson}, \citenamefont {Huang},
  \citenamefont {Dorland}, \citenamefont {Hassam}, \citenamefont {Antonsen},\
  and\ \citenamefont {Lathrop}}]{Lathrop2004}%
  \BibitemOpen
  \bibfield  {author} {\bibinfo {author} {\bibfnamefont {D.~R.}\ \bibnamefont
  {Sisan}}, \bibinfo {author} {\bibfnamefont {N.}~\bibnamefont {Mujica}},
  \bibinfo {author} {\bibfnamefont {W.~A.}\ \bibnamefont {Tillotson}}, \bibinfo
  {author} {\bibfnamefont {Y.-M.}\ \bibnamefont {Huang}}, \bibinfo {author}
  {\bibfnamefont {W.}~\bibnamefont {Dorland}}, \bibinfo {author} {\bibfnamefont
  {A.~B.}\ \bibnamefont {Hassam}}, \bibinfo {author} {\bibfnamefont {T.~M.}\
  \bibnamefont {Antonsen}},\ and\ \bibinfo {author} {\bibfnamefont {D.~P.}\
  \bibnamefont {Lathrop}},\ }\bibfield  {title} {\bibinfo {title} {Experimental
  observation and characterization of the magnetorotational instability},\
  }\href {https://doi.org/10.1103/PhysRevLett.93.114502} {\bibfield  {journal}
  {\bibinfo  {journal} {Phys. Rev. Lett.}\ }\textbf {\bibinfo {volume} {93}},\
  \bibinfo {pages} {114502} (\bibinfo {year} {2004})}\BibitemShut {NoStop}%
\bibitem [{\citenamefont {Gissinger}\ \emph {et~al.}(2011)\citenamefont
  {Gissinger}, \citenamefont {Ji},\ and\ \citenamefont
  {Goodman}}]{Gissinger2011}%
  \BibitemOpen
  \bibfield  {author} {\bibinfo {author} {\bibfnamefont {C.}~\bibnamefont
  {Gissinger}}, \bibinfo {author} {\bibfnamefont {H.}~\bibnamefont {Ji}},\ and\
  \bibinfo {author} {\bibfnamefont {J.}~\bibnamefont {Goodman}},\ }\bibfield
  {title} {\bibinfo {title} {Instabilities in magnetized spherical couette
  flow},\ }\href {https://doi.org/10.1103/PhysRevE.84.026308} {\bibfield
  {journal} {\bibinfo  {journal} {Phys. Rev. E}\ }\textbf {\bibinfo {volume}
  {84}},\ \bibinfo {pages} {026308} (\bibinfo {year} {2011})}\BibitemShut
  {NoStop}%
\bibitem [{\citenamefont {Nornberg}\ \emph {et~al.}(2010)\citenamefont
  {Nornberg}, \citenamefont {Ji}, \citenamefont {Schartman}, \citenamefont
  {Roach},\ and\ \citenamefont {Goodman}}]{Mark2010}%
  \BibitemOpen
  \bibfield  {author} {\bibinfo {author} {\bibfnamefont {M.~D.}\ \bibnamefont
  {Nornberg}}, \bibinfo {author} {\bibfnamefont {H.}~\bibnamefont {Ji}},
  \bibinfo {author} {\bibfnamefont {E.}~\bibnamefont {Schartman}}, \bibinfo
  {author} {\bibfnamefont {A.}~\bibnamefont {Roach}},\ and\ \bibinfo {author}
  {\bibfnamefont {J.}~\bibnamefont {Goodman}},\ }\bibfield  {title} {\bibinfo
  {title} {Observation of magnetocoriolis waves in a liquid metal
  taylor-couette experiment},\ }\href
  {https://doi.org/10.1103/PhysRevLett.104.074501} {\bibfield  {journal}
  {\bibinfo  {journal} {Phys. Rev. Lett.}\ }\textbf {\bibinfo {volume} {104}},\
  \bibinfo {pages} {074501} (\bibinfo {year} {2010})}\BibitemShut {NoStop}%
\bibitem [{\citenamefont {Stefani}\ \emph {et~al.}(2006)\citenamefont
  {Stefani}, \citenamefont {Gundrum}, \citenamefont {Gerbeth}, \citenamefont
  {R\"udiger}, \citenamefont {Schultz}, \citenamefont {Szklarski},\ and\
  \citenamefont {Hollerbach}}]{Stefani2006}%
  \BibitemOpen
  \bibfield  {author} {\bibinfo {author} {\bibfnamefont {F.}~\bibnamefont
  {Stefani}}, \bibinfo {author} {\bibfnamefont {T.}~\bibnamefont {Gundrum}},
  \bibinfo {author} {\bibfnamefont {G.}~\bibnamefont {Gerbeth}}, \bibinfo
  {author} {\bibfnamefont {G.}~\bibnamefont {R\"udiger}}, \bibinfo {author}
  {\bibfnamefont {M.}~\bibnamefont {Schultz}}, \bibinfo {author} {\bibfnamefont
  {J.}~\bibnamefont {Szklarski}},\ and\ \bibinfo {author} {\bibfnamefont
  {R.}~\bibnamefont {Hollerbach}},\ }\bibfield  {title} {\bibinfo {title}
  {Experimental evidence for magnetorotational instability in a taylor-couette
  flow under the influence of a helical magnetic field},\ }\href
  {https://doi.org/10.1103/PhysRevLett.97.184502} {\bibfield  {journal}
  {\bibinfo  {journal} {Phys. Rev. Lett.}\ }\textbf {\bibinfo {volume} {97}},\
  \bibinfo {pages} {184502} (\bibinfo {year} {2006})}\BibitemShut {NoStop}%
\bibitem [{\citenamefont {Liu}\ \emph {et~al.}(2006)\citenamefont {Liu},
  \citenamefont {Goodman}, \citenamefont {Herron},\ and\ \citenamefont
  {Ji}}]{Liu2006}%
  \BibitemOpen
  \bibfield  {author} {\bibinfo {author} {\bibfnamefont {W.}~\bibnamefont
  {Liu}}, \bibinfo {author} {\bibfnamefont {J.}~\bibnamefont {Goodman}},
  \bibinfo {author} {\bibfnamefont {I.}~\bibnamefont {Herron}},\ and\ \bibinfo
  {author} {\bibfnamefont {H.}~\bibnamefont {Ji}},\ }\bibfield  {title}
  {\bibinfo {title} {Helical magnetorotational instability in magnetized
  taylor-couette flow},\ }\href {https://doi.org/10.1103/PhysRevE.74.056302}
  {\bibfield  {journal} {\bibinfo  {journal} {Phys. Rev. E}\ }\textbf {\bibinfo
  {volume} {74}},\ \bibinfo {pages} {056302} (\bibinfo {year}
  {2006})}\BibitemShut {NoStop}%
\bibitem [{\citenamefont {Szklarski}\ and\ \citenamefont
  {R\"udiger}(2007)}]{Szklarski2007}%
  \BibitemOpen
  \bibfield  {author} {\bibinfo {author} {\bibfnamefont {J.}~\bibnamefont
  {Szklarski}}\ and\ \bibinfo {author} {\bibfnamefont {G.}~\bibnamefont
  {R\"udiger}},\ }\bibfield  {title} {\bibinfo {title} {Ekman-hartmann layer in
  a magnetohydrodynamic taylor-couette flow},\ }\href
  {https://doi.org/10.1103/PhysRevE.76.066308} {\bibfield  {journal} {\bibinfo
  {journal} {Phys. Rev. E}\ }\textbf {\bibinfo {volume} {76}},\ \bibinfo
  {pages} {066308} (\bibinfo {year} {2007})}\BibitemShut {NoStop}%
\bibitem [{\citenamefont {Roach}\ \emph {et~al.}(2012)\citenamefont {Roach},
  \citenamefont {Spence}, \citenamefont {Gissinger}, \citenamefont {Edlund},
  \citenamefont {Sloboda}, \citenamefont {Goodman},\ and\ \citenamefont
  {Ji}}]{Roach2012}%
  \BibitemOpen
  \bibfield  {author} {\bibinfo {author} {\bibfnamefont {A.~H.}\ \bibnamefont
  {Roach}}, \bibinfo {author} {\bibfnamefont {E.~J.}\ \bibnamefont {Spence}},
  \bibinfo {author} {\bibfnamefont {C.}~\bibnamefont {Gissinger}}, \bibinfo
  {author} {\bibfnamefont {E.~M.}\ \bibnamefont {Edlund}}, \bibinfo {author}
  {\bibfnamefont {P.}~\bibnamefont {Sloboda}}, \bibinfo {author} {\bibfnamefont
  {J.}~\bibnamefont {Goodman}},\ and\ \bibinfo {author} {\bibfnamefont
  {H.}~\bibnamefont {Ji}},\ }\bibfield  {title} {\bibinfo {title} {Observation
  of a free-shercliff-layer instability in cylindrical geometry},\ }\href
  {https://doi.org/10.1103/PhysRevLett.108.154502} {\bibfield  {journal}
  {\bibinfo  {journal} {Phys. Rev. Lett.}\ }\textbf {\bibinfo {volume} {108}},\
  \bibinfo {pages} {154502} (\bibinfo {year} {2012})}\BibitemShut {NoStop}%
\bibitem [{\citenamefont {Spence}\ \emph {et~al.}(2012)\citenamefont {Spence},
  \citenamefont {Roach}, \citenamefont {Edlund}, \citenamefont {Sloboda},\ and\
  \citenamefont {Ji}}]{Spence2012}%
  \BibitemOpen
  \bibfield  {author} {\bibinfo {author} {\bibfnamefont {E.~J.}\ \bibnamefont
  {Spence}}, \bibinfo {author} {\bibfnamefont {A.~H.}\ \bibnamefont {Roach}},
  \bibinfo {author} {\bibfnamefont {E.~M.}\ \bibnamefont {Edlund}}, \bibinfo
  {author} {\bibfnamefont {P.}~\bibnamefont {Sloboda}},\ and\ \bibinfo {author}
  {\bibfnamefont {H.}~\bibnamefont {Ji}},\ }\bibfield  {title} {\bibinfo
  {title} {Free magnetohydrodynamic shear layers in the presence of rotation
  and magnetic field},\ }\href {https://doi.org/10.1063/1.3702006} {\bibfield
  {journal} {\bibinfo  {journal} {Physics of Plasmas}\ }\textbf {\bibinfo
  {volume} {19}},\ \bibinfo {pages} {056502} (\bibinfo {year}
  {2012})}\BibitemShut {NoStop}%
\bibitem [{\citenamefont {Quataert}\ \emph {et~al.}(2002)\citenamefont
  {Quataert}, \citenamefont {Dorland},\ and\ \citenamefont
  {Hammett}}]{Quataert2002}%
  \BibitemOpen
  \bibfield  {author} {\bibinfo {author} {\bibfnamefont {E.}~\bibnamefont
  {Quataert}}, \bibinfo {author} {\bibfnamefont {W.}~\bibnamefont {Dorland}},\
  and\ \bibinfo {author} {\bibfnamefont {G.~W.}\ \bibnamefont {Hammett}},\
  }\bibfield  {title} {\bibinfo {title} {The magnetorotational instability in a
  collisionless plasma},\ }\href {https://doi.org/10.1086/342174} {\bibfield
  {journal} {\bibinfo  {journal} {The Astrophysical Journal}\ }\textbf
  {\bibinfo {volume} {577}},\ \bibinfo {pages} {524} (\bibinfo {year}
  {2002})}\BibitemShut {NoStop}%
\bibitem [{\citenamefont {Kunz}\ \emph {et~al.}(2016)\citenamefont {Kunz},
  \citenamefont {Stone},\ and\ \citenamefont {Quataert}}]{Kunz2016}%
  \BibitemOpen
  \bibfield  {author} {\bibinfo {author} {\bibfnamefont {M.~W.}\ \bibnamefont
  {Kunz}}, \bibinfo {author} {\bibfnamefont {J.~M.}\ \bibnamefont {Stone}},\
  and\ \bibinfo {author} {\bibfnamefont {E.}~\bibnamefont {Quataert}},\
  }\bibfield  {title} {\bibinfo {title} {Magnetorotational turbulence and
  dynamo in a collisionless plasma},\ }\href
  {https://doi.org/10.1103/PhysRevLett.117.235101} {\bibfield  {journal}
  {\bibinfo  {journal} {Phys. Rev. Lett.}\ }\textbf {\bibinfo {volume} {117}},\
  \bibinfo {pages} {235101} (\bibinfo {year} {2016})}\BibitemShut {NoStop}%
\bibitem [{\citenamefont {Kunz}\ \emph {et~al.}(2019)\citenamefont {Kunz},
  \citenamefont {Squire}, \citenamefont {Balbus}, \citenamefont {Bale},
  \citenamefont {Chen}, \citenamefont {Churazov}, \citenamefont {Cowley},
  \citenamefont {Forest}, \citenamefont {Gammie}, \citenamefont {Quataert},
  \citenamefont {Reynolds}, \citenamefont {Schekochihin}, \citenamefont
  {Sironi}, \citenamefont {Spitkovsky}, \citenamefont {Stone}, \citenamefont
  {Zhuravleva},\ and\ \citenamefont {Zweibel}}]{Kunz2019}%
  \BibitemOpen
  \bibfield  {author} {\bibinfo {author} {\bibfnamefont {M.~W.}\ \bibnamefont
  {Kunz}}, \bibinfo {author} {\bibfnamefont {J.}~\bibnamefont {Squire}},
  \bibinfo {author} {\bibfnamefont {S.~A.}\ \bibnamefont {Balbus}}, \bibinfo
  {author} {\bibfnamefont {S.~D.}\ \bibnamefont {Bale}}, \bibinfo {author}
  {\bibfnamefont {C.~H.~K.}\ \bibnamefont {Chen}}, \bibinfo {author}
  {\bibfnamefont {E.}~\bibnamefont {Churazov}}, \bibinfo {author}
  {\bibfnamefont {S.~C.}\ \bibnamefont {Cowley}}, \bibinfo {author}
  {\bibfnamefont {C.~B.}\ \bibnamefont {Forest}}, \bibinfo {author}
  {\bibfnamefont {C.~F.}\ \bibnamefont {Gammie}}, \bibinfo {author}
  {\bibfnamefont {E.}~\bibnamefont {Quataert}}, \bibinfo {author}
  {\bibfnamefont {C.~S.}\ \bibnamefont {Reynolds}}, \bibinfo {author}
  {\bibfnamefont {A.~A.}\ \bibnamefont {Schekochihin}}, \bibinfo {author}
  {\bibfnamefont {L.}~\bibnamefont {Sironi}}, \bibinfo {author} {\bibfnamefont
  {A.}~\bibnamefont {Spitkovsky}}, \bibinfo {author} {\bibfnamefont {J.~M.}\
  \bibnamefont {Stone}}, \bibinfo {author} {\bibfnamefont {I.}~\bibnamefont
  {Zhuravleva}},\ and\ \bibinfo {author} {\bibfnamefont {E.~G.}\ \bibnamefont
  {Zweibel}},\ }\href@noop {} {\bibinfo {title} {[plasma 2020 decadal] the
  material properties of weakly collisional, high-beta plasmas}} (\bibinfo
  {year} {2019}),\ \Eprint {https://arxiv.org/abs/1903.04080} {arXiv:1903.04080
  [physics.plasm-ph]} \BibitemShut {NoStop}%
\bibitem [{\citenamefont {Wardle}(1999)}]{wardle1999}%
  \BibitemOpen
  \bibfield  {author} {\bibinfo {author} {\bibfnamefont {M.}~\bibnamefont
  {Wardle}},\ }\bibfield  {title} {\bibinfo {title} {{The Balbus-Hawley
  instability in weakly ionized discs}},\ }\href
  {https://doi.org/10.1046/j.1365-8711.1999.02670.x} {\bibfield  {journal}
  {\bibinfo  {journal} {Monthly Notices of the Royal Astronomical Society}\
  }\textbf {\bibinfo {volume} {307}},\ \bibinfo {pages} {849} (\bibinfo {year}
  {1999})}\BibitemShut {NoStop}%
\bibitem [{\citenamefont {Wardle}\ and\ \citenamefont
  {Ng}(1999)}]{WardleNg1999}%
  \BibitemOpen
  \bibfield  {author} {\bibinfo {author} {\bibfnamefont {M.}~\bibnamefont
  {Wardle}}\ and\ \bibinfo {author} {\bibfnamefont {C.}~\bibnamefont {Ng}},\
  }\bibfield  {title} {\bibinfo {title} {{The conductivity of dense molecular
  gas}},\ }\href {https://doi.org/10.1046/j.1365-8711.1999.02211.x} {\bibfield
  {journal} {\bibinfo  {journal} {Monthly Notices of the Royal Astronomical
  Society}\ }\textbf {\bibinfo {volume} {303}},\ \bibinfo {pages} {239}
  (\bibinfo {year} {1999})}\BibitemShut {NoStop}%
\bibitem [{\citenamefont {Balbus}\ and\ \citenamefont
  {Terquem}(2001)}]{Balbus2001}%
  \BibitemOpen
  \bibfield  {author} {\bibinfo {author} {\bibfnamefont {S.~A.}\ \bibnamefont
  {Balbus}}\ and\ \bibinfo {author} {\bibfnamefont {C.}~\bibnamefont
  {Terquem}},\ }\bibfield  {title} {\bibinfo {title} {Linear analysis of the
  hall effect in protostellar disks},\ }\href {https://doi.org/10.1086/320452}
  {\bibfield  {journal} {\bibinfo  {journal} {The Astrophysical Journal}\
  }\textbf {\bibinfo {volume} {552}},\ \bibinfo {pages} {235} (\bibinfo {year}
  {2001})}\BibitemShut {NoStop}%
\bibitem [{\citenamefont {Ebrahimi}\ \emph {et~al.}(2011)\citenamefont
  {Ebrahimi}, \citenamefont {Lefebvre}, \citenamefont {Forest},\ and\
  \citenamefont {Bhattacharjee}}]{Ebrahimi2011}%
  \BibitemOpen
  \bibfield  {author} {\bibinfo {author} {\bibfnamefont {F.}~\bibnamefont
  {Ebrahimi}}, \bibinfo {author} {\bibfnamefont {B.}~\bibnamefont {Lefebvre}},
  \bibinfo {author} {\bibfnamefont {C.~B.}\ \bibnamefont {Forest}},\ and\
  \bibinfo {author} {\bibfnamefont {A.}~\bibnamefont {Bhattacharjee}},\
  }\bibfield  {title} {\bibinfo {title} {Global hall-mhd simulations of
  magnetorotational instability in a plasma couette flow experiment},\ }\href
  {https://doi.org/10.1063/1.3598481} {\bibfield  {journal} {\bibinfo
  {journal} {Physics of Plasmas}\ }\textbf {\bibinfo {volume} {18}},\ \bibinfo
  {pages} {062904} (\bibinfo {year} {2011})}\BibitemShut {NoStop}%
\bibitem [{\citenamefont {Kunz}\ and\ \citenamefont {Lesur}(2013)}]{Kunz2013}%
  \BibitemOpen
  \bibfield  {author} {\bibinfo {author} {\bibfnamefont {M.~W.}\ \bibnamefont
  {Kunz}}\ and\ \bibinfo {author} {\bibfnamefont {G.}~\bibnamefont {Lesur}},\
  }\bibfield  {title} {\bibinfo {title} {{Magnetic self-organization in
  Hall-dominated magnetorotational turbulence}},\ }\href
  {https://doi.org/10.1093/mnras/stt1171} {\bibfield  {journal} {\bibinfo
  {journal} {Monthly Notices of the Royal Astronomical Society}\ }\textbf
  {\bibinfo {volume} {434}},\ \bibinfo {pages} {2295} (\bibinfo {year}
  {2013})}\BibitemShut {NoStop}%
\bibitem [{\citenamefont {{Lesur, Geoffroy}}\ \emph {et~al.}(2014)\citenamefont
  {{Lesur, Geoffroy}}, \citenamefont {{Kunz, Matthew W.}},\ and\ \citenamefont
  {{Fromang, S\'ebastien}}}]{Lesur2014}%
  \BibitemOpen
  \bibfield  {author} {\bibinfo {author} {\bibnamefont {{Lesur, Geoffroy}}},
  \bibinfo {author} {\bibnamefont {{Kunz, Matthew W.}}},\ and\ \bibinfo
  {author} {\bibnamefont {{Fromang, S\'ebastien}}},\ }\bibfield  {title}
  {\bibinfo {title} {Thanatology in protoplanetary discs - the combined
  influence of ohmic, hall, and ambipolar diffusion on dead zones},\ }\href
  {https://doi.org/10.1051/0004-6361/201423660} {\bibfield  {journal} {\bibinfo
   {journal} {A\&A}\ }\textbf {\bibinfo {volume} {566}},\ \bibinfo {pages}
  {A56} (\bibinfo {year} {2014})}\BibitemShut {NoStop}%
\bibitem [{\citenamefont {Mininni}\ \emph {et~al.}(2002)\citenamefont
  {Mininni}, \citenamefont {G{\'{o}}mez},\ and\ \citenamefont
  {Mahajan}}]{Mininni_2002}%
  \BibitemOpen
  \bibfield  {author} {\bibinfo {author} {\bibfnamefont {P.~D.}\ \bibnamefont
  {Mininni}}, \bibinfo {author} {\bibfnamefont {D.~O.}\ \bibnamefont
  {G{\'{o}}mez}},\ and\ \bibinfo {author} {\bibfnamefont {S.~M.}\ \bibnamefont
  {Mahajan}},\ }\bibfield  {title} {\bibinfo {title} {Dynamo action in hall
  magnetohydrodynamics},\ }\href {https://doi.org/10.1086/339850} {\bibfield
  {journal} {\bibinfo  {journal} {The Astrophysical Journal}\ }\textbf
  {\bibinfo {volume} {567}},\ \bibinfo {pages} {L81} (\bibinfo {year}
  {2002})}\BibitemShut {NoStop}%
\bibitem [{\citenamefont {Mininni}\ \emph {et~al.}(2003)\citenamefont
  {Mininni}, \citenamefont {Gomez},\ and\ \citenamefont
  {Mahajan}}]{Mininni_2003}%
  \BibitemOpen
  \bibfield  {author} {\bibinfo {author} {\bibfnamefont {P.~D.}\ \bibnamefont
  {Mininni}}, \bibinfo {author} {\bibfnamefont {D.~O.}\ \bibnamefont {Gomez}},\
  and\ \bibinfo {author} {\bibfnamefont {S.~M.}\ \bibnamefont {Mahajan}},\
  }\bibfield  {title} {\bibinfo {title} {Dynamo action in magnetohydrodynamics
  and hall-magnetohydrodynamics},\ }\href {https://doi.org/10.1086/368181}
  {\bibfield  {journal} {\bibinfo  {journal} {The Astrophysical Journal}\
  }\textbf {\bibinfo {volume} {587}},\ \bibinfo {pages} {472} (\bibinfo {year}
  {2003})}\BibitemShut {NoStop}%
\bibitem [{\citenamefont {Phan}\ \emph {et~al.}(2018)\citenamefont {Phan},
  \citenamefont {Eastwood}, \citenamefont {Shay}, \citenamefont {Drake},
  \citenamefont {Sonnerup}, \citenamefont {Fujimoto}, \citenamefont {Cassak},
  \citenamefont {{\O}ieroset}, \citenamefont {Burch}, \citenamefont {Torbert}
  \emph {et~al.}}]{Phan2018}%
  \BibitemOpen
  \bibfield  {author} {\bibinfo {author} {\bibfnamefont {T.}~\bibnamefont
  {Phan}}, \bibinfo {author} {\bibfnamefont {J.}~\bibnamefont {Eastwood}},
  \bibinfo {author} {\bibfnamefont {M.}~\bibnamefont {Shay}}, \bibinfo {author}
  {\bibfnamefont {J.}~\bibnamefont {Drake}}, \bibinfo {author} {\bibfnamefont
  {B.}~\bibnamefont {Sonnerup}}, \bibinfo {author} {\bibfnamefont
  {M.}~\bibnamefont {Fujimoto}}, \bibinfo {author} {\bibfnamefont
  {P.}~\bibnamefont {Cassak}}, \bibinfo {author} {\bibfnamefont
  {M.}~\bibnamefont {{\O}ieroset}}, \bibinfo {author} {\bibfnamefont
  {J.}~\bibnamefont {Burch}}, \bibinfo {author} {\bibfnamefont
  {R.}~\bibnamefont {Torbert}}, \emph {et~al.},\ }\bibfield  {title} {\bibinfo
  {title} {Electron magnetic reconnection without ion coupling in earth?s
  turbulent magnetosheath},\ }\href@noop {} {\bibfield  {journal} {\bibinfo
  {journal} {Nature}\ }\textbf {\bibinfo {volume} {557}},\ \bibinfo {pages}
  {202} (\bibinfo {year} {2018})}\BibitemShut {NoStop}%
\bibitem [{\citenamefont {Cumming}\ \emph {et~al.}(2004)\citenamefont
  {Cumming}, \citenamefont {Arras},\ and\ \citenamefont
  {Zweibel}}]{Cumming2004}%
  \BibitemOpen
  \bibfield  {author} {\bibinfo {author} {\bibfnamefont {A.}~\bibnamefont
  {Cumming}}, \bibinfo {author} {\bibfnamefont {P.}~\bibnamefont {Arras}},\
  and\ \bibinfo {author} {\bibfnamefont {E.}~\bibnamefont {Zweibel}},\
  }\bibfield  {title} {\bibinfo {title} {Magnetic field evolution in neutron
  star crusts due to the hall effect and ohmic decay},\ }\href
  {https://doi.org/10.1086/421324} {\bibfield  {journal} {\bibinfo  {journal}
  {The Astrophysical Journal}\ }\textbf {\bibinfo {volume} {609}},\ \bibinfo
  {pages} {999} (\bibinfo {year} {2004})}\BibitemShut {NoStop}%
\bibitem [{\citenamefont {Goldreich}\ and\ \citenamefont
  {Reisenegger}(1992)}]{Goldreich1992}%
  \BibitemOpen
  \bibfield  {author} {\bibinfo {author} {\bibfnamefont {P.}~\bibnamefont
  {Goldreich}}\ and\ \bibinfo {author} {\bibfnamefont {A.}~\bibnamefont
  {Reisenegger}},\ }\bibfield  {title} {\bibinfo {title} {Magnetic field decay
  in isolated neutron stars},\ }\href@noop {} {\bibfield  {journal} {\bibinfo
  {journal} {Astrophysical Journal}\ }\textbf {\bibinfo {volume} {395}},\
  \bibinfo {pages} {250} (\bibinfo {year} {1992})}\BibitemShut {NoStop}%
\bibitem [{\citenamefont {Rheinhardt}\ and\ \citenamefont
  {Geppert}(2002)}]{Rheinhardt2002}%
  \BibitemOpen
  \bibfield  {author} {\bibinfo {author} {\bibfnamefont {M.}~\bibnamefont
  {Rheinhardt}}\ and\ \bibinfo {author} {\bibfnamefont {U.}~\bibnamefont
  {Geppert}},\ }\bibfield  {title} {\bibinfo {title} {Hall-drift induced
  magnetic field instability in neutron stars},\ }\href
  {https://doi.org/10.1103/PhysRevLett.88.101103} {\bibfield  {journal}
  {\bibinfo  {journal} {Phys. Rev. Lett.}\ }\textbf {\bibinfo {volume} {88}},\
  \bibinfo {pages} {101103} (\bibinfo {year} {2002})}\BibitemShut {NoStop}%
\bibitem [{\citenamefont {Collins}\ \emph {et~al.}(2014)\citenamefont
  {Collins}, \citenamefont {Clark}, \citenamefont {Cooper}, \citenamefont
  {Flanagan}, \citenamefont {Khalzov}, \citenamefont {Nornberg}, \citenamefont
  {Seidlitz}, \citenamefont {Wallace},\ and\ \citenamefont {Forest}}]{pcx_PoP}%
  \BibitemOpen
  \bibfield  {author} {\bibinfo {author} {\bibfnamefont {C.}~\bibnamefont
  {Collins}}, \bibinfo {author} {\bibfnamefont {M.}~\bibnamefont {Clark}},
  \bibinfo {author} {\bibfnamefont {C.~M.}\ \bibnamefont {Cooper}}, \bibinfo
  {author} {\bibfnamefont {K.}~\bibnamefont {Flanagan}}, \bibinfo {author}
  {\bibfnamefont {I.~V.}\ \bibnamefont {Khalzov}}, \bibinfo {author}
  {\bibfnamefont {M.~D.}\ \bibnamefont {Nornberg}}, \bibinfo {author}
  {\bibfnamefont {B.}~\bibnamefont {Seidlitz}}, \bibinfo {author}
  {\bibfnamefont {J.}~\bibnamefont {Wallace}},\ and\ \bibinfo {author}
  {\bibfnamefont {C.~B.}\ \bibnamefont {Forest}},\ }\bibfield  {title}
  {\bibinfo {title} {Taylor-couette flow of unmagnetized plasma},\ }\href
  {https://doi.org/10.1063/1.4872333} {\bibfield  {journal} {\bibinfo
  {journal} {Physics of Plasmas}\ }\textbf {\bibinfo {volume} {21}},\ \bibinfo
  {pages} {042117} (\bibinfo {year} {2014})}\BibitemShut {NoStop}%
\bibitem [{\citenamefont {Khalzov}\ and\ \citenamefont
  {Smolyakov}(2006)}]{Khalzov2006}%
  \BibitemOpen
  \bibfield  {author} {\bibinfo {author} {\bibfnamefont {I.}~\bibnamefont
  {Khalzov}}\ and\ \bibinfo {author} {\bibfnamefont {A.}~\bibnamefont
  {Smolyakov}},\ }\bibfield  {title} {\bibinfo {title} {On the calculation of
  steady-state magnetohydrodynamic flows of liquid metals in circular ducts of
  a rectangular cross section},\ }\href@noop {} {\bibfield  {journal} {\bibinfo
   {journal} {Technical physics}\ }\textbf {\bibinfo {volume} {51}},\ \bibinfo
  {pages} {26} (\bibinfo {year} {2006})}\BibitemShut {NoStop}%
\bibitem [{\citenamefont {Velikhov}\ \emph {et~al.}(2006)\citenamefont
  {Velikhov}, \citenamefont {Ivanov}, \citenamefont {Zakharov}, \citenamefont
  {Zakharov}, \citenamefont {Livadny},\ and\ \citenamefont
  {Serebrennikov}}]{Velikhov2006}%
  \BibitemOpen
  \bibfield  {author} {\bibinfo {author} {\bibfnamefont {E.}~\bibnamefont
  {Velikhov}}, \bibinfo {author} {\bibfnamefont {A.}~\bibnamefont {Ivanov}},
  \bibinfo {author} {\bibfnamefont {S.}~\bibnamefont {Zakharov}}, \bibinfo
  {author} {\bibfnamefont {V.}~\bibnamefont {Zakharov}}, \bibinfo {author}
  {\bibfnamefont {A.}~\bibnamefont {Livadny}},\ and\ \bibinfo {author}
  {\bibfnamefont {K.}~\bibnamefont {Serebrennikov}},\ }\bibfield  {title}
  {\bibinfo {title} {Equilibrium of current driven rotating liquid metal},\
  }\href {https://doi.org/https://doi.org/10.1016/j.physleta.2006.05.020}
  {\bibfield  {journal} {\bibinfo  {journal} {Physics Letters A}\ }\textbf
  {\bibinfo {volume} {358}},\ \bibinfo {pages} {216 } (\bibinfo {year}
  {2006})}\BibitemShut {NoStop}%
\bibitem [{\citenamefont {Khalzov}\ \emph {et~al.}(2010)\citenamefont
  {Khalzov}, \citenamefont {Smolyakov},\ and\ \citenamefont
  {Ilgisonis}}]{Khalzov2010}%
  \BibitemOpen
  \bibfield  {author} {\bibinfo {author} {\bibfnamefont {I.~V.}\ \bibnamefont
  {Khalzov}}, \bibinfo {author} {\bibfnamefont {A.~I.}\ \bibnamefont
  {Smolyakov}},\ and\ \bibinfo {author} {\bibfnamefont {V.~I.}\ \bibnamefont
  {Ilgisonis}},\ }\bibfield  {title} {\bibinfo {title} {Equilibrium
  magnetohydrodynamic flows of liquid metals in magnetorotational instability
  experiments},\ }\href {https://doi.org/10.1017/S0022112009992394} {\bibfield
  {journal} {\bibinfo  {journal} {Journal of Fluid Mechanics}\ }\textbf
  {\bibinfo {volume} {644}},\ \bibinfo {pages} {257?280} (\bibinfo {year}
  {2010})}\BibitemShut {NoStop}%
\bibitem [{\citenamefont {Weisberg}\ \emph
  {et~al.}(2017{\natexlab{a}})\citenamefont {Weisberg}, \citenamefont
  {Peterson}, \citenamefont {Milhone}, \citenamefont {Endrizzi}, \citenamefont
  {Cooper}, \citenamefont {D{\'e}sangles}, \citenamefont {Khalzov},
  \citenamefont {Siller},\ and\ \citenamefont {Forest}}]{dave_flow}%
  \BibitemOpen
  \bibfield  {author} {\bibinfo {author} {\bibfnamefont {D.}~\bibnamefont
  {Weisberg}}, \bibinfo {author} {\bibfnamefont {E.}~\bibnamefont {Peterson}},
  \bibinfo {author} {\bibfnamefont {J.}~\bibnamefont {Milhone}}, \bibinfo
  {author} {\bibfnamefont {D.}~\bibnamefont {Endrizzi}}, \bibinfo {author}
  {\bibfnamefont {C.}~\bibnamefont {Cooper}}, \bibinfo {author} {\bibfnamefont
  {V.}~\bibnamefont {D{\'e}sangles}}, \bibinfo {author} {\bibfnamefont
  {I.}~\bibnamefont {Khalzov}}, \bibinfo {author} {\bibfnamefont
  {R.}~\bibnamefont {Siller}},\ and\ \bibinfo {author} {\bibfnamefont
  {C.}~\bibnamefont {Forest}},\ }\bibfield  {title} {\bibinfo {title} {Driving
  large magnetic reynolds number flow in highly ionized, unmagnetized
  plasmas},\ }\href@noop {} {\bibfield  {journal} {\bibinfo  {journal} {Physics
  of Plasmas}\ }\textbf {\bibinfo {volume} {24}},\ \bibinfo {pages} {056502}
  (\bibinfo {year} {2017}{\natexlab{a}})}\BibitemShut {NoStop}%
\bibitem [{\citenamefont {Forest}\ \emph {et~al.}(2015)\citenamefont {Forest},
  \citenamefont {Flanagan}, \citenamefont {Brookhart}, \citenamefont {Clark},
  \citenamefont {Cooper}, \citenamefont {D{\'e}sangles}, \citenamefont
  {Egedal}, \citenamefont {Endrizzi}, \citenamefont {Khalzov}, \citenamefont
  {Li} \emph {et~al.}}]{wippl_JPP}%
  \BibitemOpen
  \bibfield  {author} {\bibinfo {author} {\bibfnamefont {C.}~\bibnamefont
  {Forest}}, \bibinfo {author} {\bibfnamefont {K.}~\bibnamefont {Flanagan}},
  \bibinfo {author} {\bibfnamefont {M.}~\bibnamefont {Brookhart}}, \bibinfo
  {author} {\bibfnamefont {M.}~\bibnamefont {Clark}}, \bibinfo {author}
  {\bibfnamefont {C.}~\bibnamefont {Cooper}}, \bibinfo {author} {\bibfnamefont
  {V.}~\bibnamefont {D{\'e}sangles}}, \bibinfo {author} {\bibfnamefont
  {J.}~\bibnamefont {Egedal}}, \bibinfo {author} {\bibfnamefont
  {D.}~\bibnamefont {Endrizzi}}, \bibinfo {author} {\bibfnamefont
  {I.}~\bibnamefont {Khalzov}}, \bibinfo {author} {\bibfnamefont
  {H.}~\bibnamefont {Li}}, \emph {et~al.},\ }\bibfield  {title} {\bibinfo
  {title} {The wisconsin plasma astrophysics laboratory},\ }\href@noop {}
  {\bibfield  {journal} {\bibinfo  {journal} {Journal of Plasma Physics}\
  }\textbf {\bibinfo {volume} {81}} (\bibinfo {year} {2015})}\BibitemShut
  {NoStop}%
\bibitem [{\citenamefont {Cooper}\ \emph {et~al.}(2014)\citenamefont {Cooper},
  \citenamefont {Wallace}, \citenamefont {Brookhart}, \citenamefont {Clark},
  \citenamefont {Collins}, \citenamefont {Ding}, \citenamefont {Flanagan},
  \citenamefont {Khalzov}, \citenamefont {Li}, \citenamefont {Milhone} \emph
  {et~al.}}]{mpdx_PoP}%
  \BibitemOpen
  \bibfield  {author} {\bibinfo {author} {\bibfnamefont {C.}~\bibnamefont
  {Cooper}}, \bibinfo {author} {\bibfnamefont {J.}~\bibnamefont {Wallace}},
  \bibinfo {author} {\bibfnamefont {M.}~\bibnamefont {Brookhart}}, \bibinfo
  {author} {\bibfnamefont {M.}~\bibnamefont {Clark}}, \bibinfo {author}
  {\bibfnamefont {C.}~\bibnamefont {Collins}}, \bibinfo {author} {\bibfnamefont
  {W.}~\bibnamefont {Ding}}, \bibinfo {author} {\bibfnamefont {K.}~\bibnamefont
  {Flanagan}}, \bibinfo {author} {\bibfnamefont {I.}~\bibnamefont {Khalzov}},
  \bibinfo {author} {\bibfnamefont {Y.}~\bibnamefont {Li}}, \bibinfo {author}
  {\bibfnamefont {J.}~\bibnamefont {Milhone}}, \emph {et~al.},\ }\bibfield
  {title} {\bibinfo {title} {The madison plasma dynamo experiment: A facility
  for studying laboratory plasma astrophysics},\ }\href@noop {} {\bibfield
  {journal} {\bibinfo  {journal} {Physics of Plasmas}\ }\textbf {\bibinfo
  {volume} {21}},\ \bibinfo {pages} {013505} (\bibinfo {year}
  {2014})}\BibitemShut {NoStop}%
\bibitem [{\citenamefont {Weisberg}(2016)}]{DaveThesis}%
  \BibitemOpen
  \bibfield  {author} {\bibinfo {author} {\bibfnamefont {D.~B.}\ \bibnamefont
  {Weisberg}},\ }\emph {\bibinfo {title} {Pursuing the plasma dynamo and MRI in
  the laboratory}},\ \href@noop {} {Ph.D. thesis},\ \bibinfo  {school}
  {University of Wisconsin-Madison} (\bibinfo {year} {2016})\BibitemShut
  {NoStop}%
\bibitem [{\citenamefont {Weisberg}\ \emph
  {et~al.}(2017{\natexlab{b}})\citenamefont {Weisberg}, \citenamefont
  {Peterson}, \citenamefont {Milhone}, \citenamefont {Endrizzi}, \citenamefont
  {Cooper}, \citenamefont {D\'esangles}, \citenamefont {Khalzov}, \citenamefont
  {Siller},\ and\ \citenamefont {Forest}}]{DavePoPFlow}%
  \BibitemOpen
  \bibfield  {author} {\bibinfo {author} {\bibfnamefont {D.~B.}\ \bibnamefont
  {Weisberg}}, \bibinfo {author} {\bibfnamefont {E.}~\bibnamefont {Peterson}},
  \bibinfo {author} {\bibfnamefont {J.}~\bibnamefont {Milhone}}, \bibinfo
  {author} {\bibfnamefont {D.}~\bibnamefont {Endrizzi}}, \bibinfo {author}
  {\bibfnamefont {C.}~\bibnamefont {Cooper}}, \bibinfo {author} {\bibfnamefont
  {V.}~\bibnamefont {D\'esangles}}, \bibinfo {author} {\bibfnamefont
  {I.}~\bibnamefont {Khalzov}}, \bibinfo {author} {\bibfnamefont
  {R.}~\bibnamefont {Siller}},\ and\ \bibinfo {author} {\bibfnamefont {C.~B.}\
  \bibnamefont {Forest}},\ }\bibfield  {title} {\bibinfo {title} {Driving large
  magnetic reynolds number flow in highly ionized, unmagnetized plasmas},\
  }\href {https://doi.org/10.1063/1.4978889} {\bibfield  {journal} {\bibinfo
  {journal} {Physics of Plasmas}\ }\textbf {\bibinfo {volume} {24}},\ \bibinfo
  {pages} {056502} (\bibinfo {year} {2017}{\natexlab{b}})}\BibitemShut
  {NoStop}%
\bibitem [{\citenamefont {Peterson}(2019)}]{Ethan_Thesis}%
  \BibitemOpen
  \bibfield  {author} {\bibinfo {author} {\bibfnamefont {E.~E.}\ \bibnamefont
  {Peterson}},\ }\emph {\bibinfo {title} {A Laboratory Model for Magnetized
  Stellar Winds}},\ \href@noop {} {Ph.D. thesis},\ \bibinfo  {school}
  {University of Wisconsin-Madison} (\bibinfo {year} {2019})\BibitemShut
  {NoStop}%
\bibitem [{\citenamefont {Milhone}\ \emph {et~al.}(2019)\citenamefont
  {Milhone}, \citenamefont {Flanagan}, \citenamefont {Nornberg}, \citenamefont
  {Tabbutt},\ and\ \citenamefont {Forest}}]{jason_rsi}%
  \BibitemOpen
  \bibfield  {author} {\bibinfo {author} {\bibfnamefont {J.}~\bibnamefont
  {Milhone}}, \bibinfo {author} {\bibfnamefont {K.}~\bibnamefont {Flanagan}},
  \bibinfo {author} {\bibfnamefont {M.~D.}\ \bibnamefont {Nornberg}}, \bibinfo
  {author} {\bibfnamefont {M.}~\bibnamefont {Tabbutt}},\ and\ \bibinfo {author}
  {\bibfnamefont {C.~B.}\ \bibnamefont {Forest}},\ }\bibfield  {title}
  {\bibinfo {title} {A spectrometer for high-precision ion temperature and
  velocity measurements in low-temperature plasmas},\ }\href@noop {} {\bibfield
   {journal} {\bibinfo  {journal} {Review of Scientific Instruments}\ }\textbf
  {\bibinfo {volume} {90}},\ \bibinfo {pages} {063502} (\bibinfo {year}
  {2019})}\BibitemShut {NoStop}%
\bibitem [{\citenamefont {Rayleigh}(1879)}]{Rayleigh1879}%
  \BibitemOpen
  \bibfield  {author} {\bibinfo {author} {\bibfnamefont {L.}~\bibnamefont
  {Rayleigh}},\ }\bibfield  {title} {\bibinfo {title} {On the stability, or
  instability, of certain fluid motions},\ }\href
  {https://doi.org/10.1112/plms/s1-11.1.57} {\bibfield  {journal} {\bibinfo
  {journal} {Proceedings of the London Mathematical Society}\ }\textbf
  {\bibinfo {volume} {s1-11}},\ \bibinfo {pages} {57} (\bibinfo {year}
  {1879})}\BibitemShut {NoStop}%
\bibitem [{\citenamefont {Balbus}\ and\ \citenamefont
  {Hawley}(1991)}]{balbus1991}%
  \BibitemOpen
  \bibfield  {author} {\bibinfo {author} {\bibfnamefont {S.~A.}\ \bibnamefont
  {Balbus}}\ and\ \bibinfo {author} {\bibfnamefont {J.~F.}\ \bibnamefont
  {Hawley}},\ }\bibfield  {title} {\bibinfo {title} {A powerful local shear
  instability in weakly magnetized disks. i-linear analysis. ii-nonlinear
  evolution},\ }\href@noop {} {\bibfield  {journal} {\bibinfo  {journal} {The
  Astrophysical Journal}\ }\textbf {\bibinfo {volume} {376}},\ \bibinfo {pages}
  {214} (\bibinfo {year} {1991})}\BibitemShut {NoStop}%
\bibitem [{\citenamefont {Cooper}\ \emph {et~al.}(2016)\citenamefont {Cooper},
  \citenamefont {Weisberg}, \citenamefont {Khalzov}, \citenamefont {Milhone},
  \citenamefont {Flanagan}, \citenamefont {Peterson}, \citenamefont {Wahl},\
  and\ \citenamefont {Forest}}]{Cooper2016_cusp}%
  \BibitemOpen
  \bibfield  {author} {\bibinfo {author} {\bibfnamefont {C.~M.}\ \bibnamefont
  {Cooper}}, \bibinfo {author} {\bibfnamefont {D.~B.}\ \bibnamefont
  {Weisberg}}, \bibinfo {author} {\bibfnamefont {I.}~\bibnamefont {Khalzov}},
  \bibinfo {author} {\bibfnamefont {J.}~\bibnamefont {Milhone}}, \bibinfo
  {author} {\bibfnamefont {K.}~\bibnamefont {Flanagan}}, \bibinfo {author}
  {\bibfnamefont {E.}~\bibnamefont {Peterson}}, \bibinfo {author}
  {\bibfnamefont {C.}~\bibnamefont {Wahl}},\ and\ \bibinfo {author}
  {\bibfnamefont {C.~B.}\ \bibnamefont {Forest}},\ }\bibfield  {title}
  {\bibinfo {title} {Direct measurement of the plasma loss width in an
  optimized, high ionization fraction, magnetic multi-dipole ring cusp},\
  }\href {https://doi.org/10.1063/1.4963850} {\bibfield  {journal} {\bibinfo
  {journal} {Physics of Plasmas}\ }\textbf {\bibinfo {volume} {23}},\ \bibinfo
  {pages} {102505} (\bibinfo {year} {2016})}\BibitemShut {NoStop}%
\bibitem [{\citenamefont {Sovinec}\ \emph {et~al.}(2004)\citenamefont
  {Sovinec}, \citenamefont {Glasser}, \citenamefont {Gianakon}, \citenamefont
  {Barnes}, \citenamefont {Nebel}, \citenamefont {Kruger}, \citenamefont
  {Plimpton}, \citenamefont {Tarditi}, \citenamefont {Chu},\ and\ \citenamefont
  {the NIMROD~Team}}]{nimrod}%
  \BibitemOpen
  \bibfield  {author} {\bibinfo {author} {\bibfnamefont {C.}~\bibnamefont
  {Sovinec}}, \bibinfo {author} {\bibfnamefont {A.}~\bibnamefont {Glasser}},
  \bibinfo {author} {\bibfnamefont {T.}~\bibnamefont {Gianakon}}, \bibinfo
  {author} {\bibfnamefont {D.}~\bibnamefont {Barnes}}, \bibinfo {author}
  {\bibfnamefont {R.}~\bibnamefont {Nebel}}, \bibinfo {author} {\bibfnamefont
  {S.}~\bibnamefont {Kruger}}, \bibinfo {author} {\bibfnamefont
  {S.}~\bibnamefont {Plimpton}}, \bibinfo {author} {\bibfnamefont
  {A.}~\bibnamefont {Tarditi}}, \bibinfo {author} {\bibfnamefont
  {M.}~\bibnamefont {Chu}},\ and\ \bibinfo {author} {\bibnamefont {the
  NIMROD~Team}},\ }\bibfield  {title} {\bibinfo {title} {Nonlinear
  magnetohydrodynamics with high-order finite elements},\ }\href@noop {}
  {\bibfield  {journal} {\bibinfo  {journal} {J. Comp. Phys.}\ }\textbf
  {\bibinfo {volume} {195}},\ \bibinfo {pages} {355} (\bibinfo {year}
  {2004})}\BibitemShut {NoStop}%
\bibitem [{\citenamefont {Peterson}\ \emph {et~al.}(2019)\citenamefont
  {Peterson}, \citenamefont {Endrizzi}, \citenamefont {Beidler}, \citenamefont
  {Bunkers}, \citenamefont {Clark}, \citenamefont {Egedal}, \citenamefont
  {Flanagan}, \citenamefont {McCollam}, \citenamefont {Milhone}, \citenamefont
  {Olson} \emph {et~al.}}]{Ethan_Nature}%
  \BibitemOpen
  \bibfield  {author} {\bibinfo {author} {\bibfnamefont {E.~E.}\ \bibnamefont
  {Peterson}}, \bibinfo {author} {\bibfnamefont {D.~A.}\ \bibnamefont
  {Endrizzi}}, \bibinfo {author} {\bibfnamefont {M.}~\bibnamefont {Beidler}},
  \bibinfo {author} {\bibfnamefont {K.~J.}\ \bibnamefont {Bunkers}}, \bibinfo
  {author} {\bibfnamefont {M.}~\bibnamefont {Clark}}, \bibinfo {author}
  {\bibfnamefont {J.}~\bibnamefont {Egedal}}, \bibinfo {author} {\bibfnamefont
  {K.}~\bibnamefont {Flanagan}}, \bibinfo {author} {\bibfnamefont {K.~J.}\
  \bibnamefont {McCollam}}, \bibinfo {author} {\bibfnamefont {J.}~\bibnamefont
  {Milhone}}, \bibinfo {author} {\bibfnamefont {J.}~\bibnamefont {Olson}},
  \emph {et~al.},\ }\bibfield  {title} {\bibinfo {title} {A laboratory model
  for the parker spiral and magnetized stellar winds},\ }\href@noop {}
  {\bibfield  {journal} {\bibinfo  {journal} {Nature Physics}\ }\textbf
  {\bibinfo {volume} {15}},\ \bibinfo {pages} {1095} (\bibinfo {year}
  {2019})}\BibitemShut {NoStop}%
\bibitem [{\citenamefont {Stenzel}\ and\ \citenamefont
  {Urrutia}(2000)}]{Stenzel2000}%
  \BibitemOpen
  \bibfield  {author} {\bibinfo {author} {\bibfnamefont {R.~L.}\ \bibnamefont
  {Stenzel}}\ and\ \bibinfo {author} {\bibfnamefont {J.~M.}\ \bibnamefont
  {Urrutia}},\ }\bibfield  {title} {\bibinfo {title} {Electron
  magnetohydrodynamic turbulence in a high-beta plasma. i. plasma parameters
  and instability conditions},\ }\href {https://doi.org/10.1063/1.1314343}
  {\bibfield  {journal} {\bibinfo  {journal} {Physics of Plasmas}\ }\textbf
  {\bibinfo {volume} {7}},\ \bibinfo {pages} {4450} (\bibinfo {year}
  {2000})}\BibitemShut {NoStop}%
\bibitem [{\citenamefont {Birn}\ \emph {et~al.}(2001)\citenamefont {Birn},
  \citenamefont {Drake}, \citenamefont {Shay}, \citenamefont {Rogers},
  \citenamefont {Denton}, \citenamefont {Hesse}, \citenamefont {Kuznetsova},
  \citenamefont {Ma}, \citenamefont {Bhattacharjee}, \citenamefont {Otto},\
  and\ \citenamefont {Pritchett}}]{Birn2001}%
  \BibitemOpen
  \bibfield  {author} {\bibinfo {author} {\bibfnamefont {J.}~\bibnamefont
  {Birn}}, \bibinfo {author} {\bibfnamefont {J.~F.}\ \bibnamefont {Drake}},
  \bibinfo {author} {\bibfnamefont {M.~A.}\ \bibnamefont {Shay}}, \bibinfo
  {author} {\bibfnamefont {B.~N.}\ \bibnamefont {Rogers}}, \bibinfo {author}
  {\bibfnamefont {R.~E.}\ \bibnamefont {Denton}}, \bibinfo {author}
  {\bibfnamefont {M.}~\bibnamefont {Hesse}}, \bibinfo {author} {\bibfnamefont
  {M.}~\bibnamefont {Kuznetsova}}, \bibinfo {author} {\bibfnamefont {Z.~W.}\
  \bibnamefont {Ma}}, \bibinfo {author} {\bibfnamefont {A.}~\bibnamefont
  {Bhattacharjee}}, \bibinfo {author} {\bibfnamefont {A.}~\bibnamefont
  {Otto}},\ and\ \bibinfo {author} {\bibfnamefont {P.~L.}\ \bibnamefont
  {Pritchett}},\ }\bibfield  {title} {\bibinfo {title} {Geospace environmental
  modeling (gem) magnetic reconnection challenge},\ }\href
  {https://doi.org/10.1029/1999JA900449} {\bibfield  {journal} {\bibinfo
  {journal} {Journal of Geophysical Research: Space Physics}\ }\textbf
  {\bibinfo {volume} {106}},\ \bibinfo {pages} {3715} (\bibinfo {year}
  {2001})}\BibitemShut {NoStop}%
\bibitem [{\citenamefont {Olson}\ \emph {et~al.}(2016)\citenamefont {Olson},
  \citenamefont {Egedal}, \citenamefont {Greess}, \citenamefont {Myers},
  \citenamefont {Clark}, \citenamefont {Endrizzi}, \citenamefont {Flanagan},
  \citenamefont {Milhone}, \citenamefont {Peterson}, \citenamefont {Wallace},
  \citenamefont {Weisberg},\ and\ \citenamefont {Forest}}]{JoePRL}%
  \BibitemOpen
  \bibfield  {author} {\bibinfo {author} {\bibfnamefont {J.}~\bibnamefont
  {Olson}}, \bibinfo {author} {\bibfnamefont {J.}~\bibnamefont {Egedal}},
  \bibinfo {author} {\bibfnamefont {S.}~\bibnamefont {Greess}}, \bibinfo
  {author} {\bibfnamefont {R.}~\bibnamefont {Myers}}, \bibinfo {author}
  {\bibfnamefont {M.}~\bibnamefont {Clark}}, \bibinfo {author} {\bibfnamefont
  {D.}~\bibnamefont {Endrizzi}}, \bibinfo {author} {\bibfnamefont
  {K.}~\bibnamefont {Flanagan}}, \bibinfo {author} {\bibfnamefont
  {J.}~\bibnamefont {Milhone}}, \bibinfo {author} {\bibfnamefont
  {E.}~\bibnamefont {Peterson}}, \bibinfo {author} {\bibfnamefont
  {J.}~\bibnamefont {Wallace}}, \bibinfo {author} {\bibfnamefont
  {D.}~\bibnamefont {Weisberg}},\ and\ \bibinfo {author} {\bibfnamefont
  {C.~B.}\ \bibnamefont {Forest}},\ }\bibfield  {title} {\bibinfo {title}
  {Experimental demonstration of the collisionless plasmoid instability below
  the ion kinetic scale during magnetic reconnection},\ }\href
  {https://doi.org/10.1103/PhysRevLett.116.255001} {\bibfield  {journal}
  {\bibinfo  {journal} {Phys. Rev. Lett.}\ }\textbf {\bibinfo {volume} {116}},\
  \bibinfo {pages} {255001} (\bibinfo {year} {2016})}\BibitemShut {NoStop}%
\bibitem [{\citenamefont {Ren}\ \emph {et~al.}(2005)\citenamefont {Ren},
  \citenamefont {Yamada}, \citenamefont {Gerhardt}, \citenamefont {Ji},
  \citenamefont {Kulsrud},\ and\ \citenamefont {Kuritsyn}}]{MRX_Hall}%
  \BibitemOpen
  \bibfield  {author} {\bibinfo {author} {\bibfnamefont {Y.}~\bibnamefont
  {Ren}}, \bibinfo {author} {\bibfnamefont {M.}~\bibnamefont {Yamada}},
  \bibinfo {author} {\bibfnamefont {S.}~\bibnamefont {Gerhardt}}, \bibinfo
  {author} {\bibfnamefont {H.}~\bibnamefont {Ji}}, \bibinfo {author}
  {\bibfnamefont {R.}~\bibnamefont {Kulsrud}},\ and\ \bibinfo {author}
  {\bibfnamefont {A.}~\bibnamefont {Kuritsyn}},\ }\bibfield  {title} {\bibinfo
  {title} {Experimental verification of the hall effect during magnetic
  reconnection in a laboratory plasma},\ }\href
  {https://doi.org/10.1103/PhysRevLett.95.055003} {\bibfield  {journal}
  {\bibinfo  {journal} {Phys. Rev. Lett.}\ }\textbf {\bibinfo {volume} {95}},\
  \bibinfo {pages} {055003} (\bibinfo {year} {2005})}\BibitemShut {NoStop}%
\end{thebibliography}
\end{document}